\documentclass[sigconf,nonacm]{acmart}

\usepackage{xcolor}
\usepackage{tikz}
\usepackage{amsmath}
\usepackage{xspace}
\usepackage{siunitx}
\usepackage{subcaption}
\usepackage{caption}
\usepackage{float}
\usepackage{paralist}
\usepackage{balance}
\usepackage{ulem}
\usepackage{titlesec}
\usepackage{hhline}
\usepackage{tabularx}
\usepackage[title]{appendix}

\usepackage[
type={CC},
modifier={by},
version={4.0},
]{doclicense}
\usepackage{etoolbox}

\newcommand\blfootnote[1]{%
  \begingroup
  \renewcommand\thefootnote{}\footnote{#1}%
  \addtocounter{footnote}{-1}%
  \endgroup
}

\titlespacing\section{0pt}{8pt plus 2pt minus 0pt}{0pt plus 2pt minus 2pt}
\titlespacing\subsection{0pt}{6pt plus 2pt minus 2pt}{0pt plus 2pt minus 2pt}
\titlespacing\subsubsection{0pt}{6pt plus 2pt minus 2pt}{0pt plus 2pt minus 2pt}
\begin{document}

\def\sqp{{SQP}\xspace}
\def\bbr{{BBR}\xspace}
\def\copa{{Copa}\xspace}
\def\sprout{{Sprout}\xspace}
\def\pcc{{PCC}\xspace}
\def\vivace{{Vivace}\xspace}
\def\sprout{{Sprout}\xspace}
\def\vegas{{Vegas}\xspace}
\def\cubic{{Cubic}\xspace}
\def\webrtc{{WebRTC}\xspace}
\def\company{{Google}\xspace}

\newcommand{\revised}[2]{{\color{cyan}[Revised: XXXX #1 XXXX]}{\color{olive}[New: #2]}}
\newcommand{\todo}[1]{{\color{red}[TODO: #1]}}
\newcommand{\note}[1]{{\color{orange}[Note: #1]}}
\newcommand{\dd}[1]{{\color{orange}[Devdeep: #1]}}
\newcommand{\connor}[1]{{\color{red}[Connor: #1]}}
\newcommand{\teng}[1]{{\color{blue}[Teng: #1]}}
\newcommand{\srini}[1]{{\color{magenta}[Srini: #1]}}
\newcommand{\david}[1]{{\color{magenta}[David: #1]}}
\newcommand{\neal}[1]{{\color{red}[Neal: #1]}}
\newcommand{\rewrite}[1]{{\color{red}[Rewrite: \color{brown}#1\color{red}]}}


\renewcommand{\revised}[2]{{\color{olive}#2}}

\renewcommand{\revised}[2]{#2}

 \renewcommand{\dd}[1]{}
 \renewcommand{\connor}[1]{}
 \renewcommand{\teng}[1]{}
 \renewcommand{\srini}[1]{}
 \renewcommand{\david}[1]{}
 \renewcommand{\neal}[1]{}

 \renewcommand{\note}[1]{}
 
\settopmatter{printfolios=true}

\title{SQP: Congestion Control for Low-Latency Interactive Video Streaming}

\author{Devdeep Ray}
\affiliation{\institution{Carnegie Mellon University, Google}
\country{ }}
\author{Connor Smith}
\affiliation{\institution{Google}
\country{ }}
\author{Teng Wei}
\affiliation{\institution{Google}
\country{ }}
\author{David Chu}
\affiliation{\institution{Google}
\country{ }}
\author{Srinivasan Seshan}
\affiliation{\institution{Google}
\country{ }}

\renewcommand{\shortauthors}{Devdeep Ray, Connor Smith, Teng Wei, David Chu, \& Srinivasan Seshan}


 \begin{abstract}

\revised{
This paper presents the design and evaluation of \sqp \footnote{Streaming Quality Protocol, Snoqualmie Pass}, a congestion control algorithm (CCA) for low-latency interactive video streaming applications like AR streaming and cloud gaming.
\sqp couples network measurements with frame transmissions, and responds to congestion primarily via modulation of the video bitrate. 
\sqp's tight integration with the traffic pattern of interactive video streaming also enables a unique rate- and delay-based approach for measuring the network bandwidth. This combination of features enables \sqp to perform better than prior video-agnostic CCA designs by  minimizing end-to-end frame delay due to sender-side queuing, and achieving low network queuing delay, while remaining competitive in the presence of queue-building cross traffic.
}
{
This paper presents the design and evaluation of \sqp \footnote{Streaming Quality Protocol, Snoqualmie Pass}, a congestion control algorithm (CCA) for interactive video streaming applications that need to stream high-bitrate compressed video with very low end-to-end frame delay (eg. AR streaming, cloud gaming). 
\sqp uses frame-coupled, paced packet trains to sample the network bandwidth, and uses an adaptive one-way delay measurement to recover from queuing, for low, bounded queuing delay.
\sqp rapidly adapts to changes in the link bandwidth, ensuring high utilization and low frame delay, and also achieves competitive bandwidth shares when competing with queue-building flows within an acceptable delay envelope.
\sqp has good fairness properties, and works well on links with shallow buffers.
}

In real-world A/B testing of \sqp against \copa in \company's AR streaming platform, \sqp achieves 27\% and 15\% more sessions with high bitrate and low frame delay for LTE and Wi-Fi, respectively.
\revised{
On emulated wireless network traces, \sqp achieves $\approx 2\times$ higher throughput compared to GoogCC (WebRTC), and $\approx 1.7-4 \times$ lower end-to-end frame delay compared to \copa (with mode switching), Sprout and BBR.
}{}
When competing with queue-building traffic like \cubic and \bbr, \sqp achieves $2-3\times$ higher bandwidth compared to GoogCC (\webrtc), \sprout, and PCC-\vivace, and comparable performance to Copa (with mode switching).

\end{abstract}

\maketitle

\blfootnote{\begin{minipage}{0.3\columnwidth}
     \href{https://creativecommons.org/licenses/by/4.0/}{\includegraphics[width=0.90\textwidth]{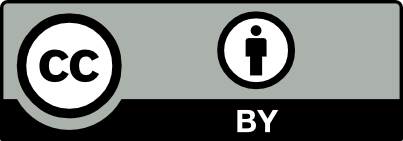}}
  \end{minipage}
  \begin{minipage}{0.6\columnwidth}
     \href{https://creativecommons.org/licenses/by/4.0/}{This work is licensed under a Creative Commons Attribution International 4.0 License.}
  \end{minipage}

  \vspace{0.1in}
  }

\section{Introduction}
\label{sec:introduction}
In recent years, there has been an increasing interest in deploying a new class of video streaming applications: low-latency, interactive video streaming~\cite{cloud-survey}.
These applications offload demanding graphics-intensive workloads like video games and 3D model rendering to powerful cloud servers at the edge, and stream the application view-port to clients over the Internet in the form of compressed video.
Examples of deployed applications that use this approach include cloud gaming services (e.g., Amazon Luna~\cite{amz-luna}, Google Stadia~\cite{stadia}, Microsoft xCloud~\cite{xcloud}, and NVIDIA GeForce Now~\cite{geforcenow}), remote desktop services (e.g., Azure Virtual Desktop~\cite{azurevirtualdesktop}, Chrome Remote Desktop~\cite{crdp}, and others),
and cloud augmented reality services (e.g., 3D car search on Android~\cite{google-ar-stream}, ARR~\cite{azure-remote-rendering}, NVIDIA CloudXR~\cite{nvidia-cloudxr}, ) that enable high quality augmented reality (AR) on mobile devices.

\revised{
Low-latency interactive streaming applications have unique properties: video is encoded in real-time according to a bitrate chosen based on current network conditions.
In order to provide end users a comparable experience to running the applications locally, video should be encoded at the highest bitrate possible without increasing the end-to-end frame transmission delay.
Congestion control mechanisms used in existing low-latency video streaming applications can be categorized into two broad approaches:

\begin{compactenum}
    \item \emph{Decoupled CCA/Video} : One simple way to stream interactive content is to use a conventional (video-agnostic) CCA in conjunction with an adaptive bitrate algorithm (e.g., HLS~\cite{rfc8216hls}, MPEG-DASH~\cite{sodagar2011mpeg}). However, conventional CCAs are designed to optimize throughput and \textit{per-packet} delay for an infinite backlog of data, and perform sub-optimally when evaluated for \textit{end-to-end video frame} delay. Fundamentally, this is caused due to the loose coupling between the encoder and the CCA, resulting in a mismatch between the instantaneous CCA transmission rate and the size of encoded video frames.
    \item \emph{Coupled CCA/Video} : Salsify~\cite{fouladi2018salsify} and WebRTC~\cite{carlucci2016analysis} achieve improved streaming performance by coupling the network and encoder layers: Salsify enables the encoder to make just-in-time bitrate decisions for each frame, whereas \webrtc matches the video bitrate to 1-second average network capacity estimates.
\end{compactenum}
}
{

For the end user experience to be comparable to running the applications locally, the video must be encoded at the highest bitrate that still allows the frames to be transmitted and received with minimal delay.
A CCA for low-latency interactive video streaming must have the following properties:
\begin{compactenum}
    \item \emph{Low Queuing Delay: } The CCA must be able to probe for more bandwidth without causing excessive queuing, and must quickly back off when the available bandwidth decreases in order to reduce in-network queuing. CCAs like \cubic~\cite{ha2008cubic} fill up network queues until packet loss occurs, and some CCAs, like \pcc~\cite{dong2015pcc}, are slow to react to drops in bandwidth, resulting in very high delays that are unacceptable for low-latency interactive streaming.
    
    \item \emph{Fairness: } 
    The CCA must achieve high, stable bandwidth when competing with queue-building flows (e.g., \cubic, \bbr~\cite{cardwell2016bbr}), while achieving low delay when running in isolation. Some low-delay CCAs have explicit mechanisms to prioritize throughput over delay when queue-building cross traffic is detected, but they can be inherently unstable (e.g., Copa~\cite{arun2018copa} can misdetect self-induced queuing as competing traffic, resulting in additional self-induced queuing~\cite{goyal2018elasticity-nimbus}), while others are slow to converge (e.g., Nimbus~\cite{goyal2018elasticity-nimbus} operates over 10s of seconds).  In addition, multiple homogeneous flows must also converge to fairness quickly.
    \item \emph{Friendliness: } The CCA must be friendly to other CCAs and must avoid starving them - GoogCC's~\cite{ietf-rmcat-gcc-02,carlucci2016analysis,webrtc-eval} adaptive threshold mechanism for competing with cross traffic can occasionally starve other flows (e.g., \cubic).
    \item \emph{Video Awareness: } The CCA must accommodate encoder frame size variations, and achieve bandwidth probing in application-limited scenarios without the need for frame padding. 
    The CCA must use a rate-based congestion control mechanism to minimize the end-to-end frame delay - the bursty nature and time-varying throughput of window-based mechanisms necessitate an undesirable trade-off between bandwidth utilization, encoder rate-control updates, and the end-to-end frame delay.
\end{compactenum}


While most CCAs aim to achieve high throughput, low delay, and competitive performance when competing with queue-building flows, simultaneously achieving these requirements is challenging in an environment as diverse as the Internet.
Choosing the right trade-offs and correctly prioritizing the design requirements (listed above in decreasing order of priority) enables a design that is highly optimized for the specific application class.
Existing CCAs make different trade-offs based on their particular design goals, and some of these design choices make them unsuitable for low-latency interactive streaming applications.

}

\revised{
In this paper, we present \sqp, a novel congestion control algorithm that was developed in conjunction with \company's AR streaming platform and is suitable for low-latency interactive streaming applications like cloud gaming and cloud AR. \sqp's design extends the level of CCA-codec integration seen in systems like Salsify and WebRTC by tightly coupling network measurements with frame transmissions in addition to coupling the CCA and the choice of the video encoding bitrate.
This approach enables \sqp to meet the stringent frame-delay requirements of low-latency interactive streaming applications.
It also enables \sqp to achieve higher throughput than existing low delay algorithms like \sprout (Salsify) and GoogCC~\cite{ietf-rmcat-gcc-02} (\webrtc) on dynamic links, and when competing with queue-building cross traffic.

\sqp's key features are listed below: 
\begin{compactenum}
    \item \emph{Integrated Bandwidth Sampling} : SQP leverages the frame transmission traffic pattern of interactive streaming to sample the network bandwidth without increasing the end-to-end frame delay due to queuing, in the common case when the available bandwidth is stable.
    \item \emph{Direct Video Bitrate Control} : SQP performs congestion control by directly modulating the video encoder bitrate according to the measured network capacity.
    \item \emph{Competitive Bandwidth Estimates} : SQP's hybrid rate- and delay-based bandwidth estimation mechanism enables \sqp to achieve a competitive throughput share by default in the presence of queue-building cross traffic, while simultaneously achieving very low end-to-end frame delay in isolation.
    While Copa~\cite{arun2018copa} switches to competitive mode when queuing delays increase, \sqp has a mode-free bandwidth adaptation algorithm that is able to avoid persistent self-induced queuing, but is competitive when cross traffic creates a standing queue.
    \item \emph{Friendly to Cross Traffic}: \sqp's frame pacing and target multiplier mechanisms provide a theoretical upper bound on its link share when competing with cross traffic, ensuring friendliness to other congestion control algorithms.
\end{compactenum}
}
{

In this paper, we present \sqp, a novel congestion control algorithm that was developed in conjunction with \company's AR streaming platform.
\sqp's key features are listed below:

\begin{compactenum}
    \item \emph{Prioritizing Delay over Link Utilization: } Since delay is more critical for the QoE of low-latency interactive video streaming applications, \sqp sacrifices peak bandwidth utilization when running in isolation in order to achieve low delay and delay stability. For example, on a 20 Mbps link where \sqp is the only flow, it is acceptable to utilize 18 Mbps if this trade-off reduces delays across a wider range of scenarios.
    \item \emph{Application-specific Trade-offs :} \sqp is designed for low-latency interactive streaming applications, which have specific requirements in terms of minimum bandwidth and maximum delay. If these parameters are outside the acceptable range due to external factors (e.g., poor link conditions, very high delays due to queue-building cross traffic), it is acceptable to end the streaming session. In contrast to traditional algorithms, \sqp restricts its operating environment, which enables \sqp to achieve acceptable throughput and delay performance across a wider range of relevant scenarios.
    \item \emph{Frame-focused Operation :}  In-network queuing is a key mechanism that allows CCAs to detect the network capacity. CCAs that probe infrequently (e.g., \pcc, GoogCC) have lower average delay, but suffer from link underutilization on variable links. \sqp piggy-backs bandwidth measurements onto each frame's transmission by sending each frame as a short (paced) burst, and updates its bandwidth estimate after receiving feedback for each frame. For low-latency interactive streaming applications, the QoE is determined by the end-to-end frame delay, and not just the in-network queuing delay. \sqp network probing relies on queuing at the sub-frame level without increasing the end-to-end frame delay, and is able to adapt to changes in network bandwidth much faster than protocols like PCC~\cite{dong2015pcc, dong2018pcc} and GoogCC~\cite{carlucci2016analysis}.
    \item \emph{Direct Video Bitrate Control :} \sqp uses frame-level bitrate changes in order to respond to congestion, and drains self-induced queues by reducing the video bitrate.
    \sqp's rate-based congestion control minimizes the end-to-end frame delay compared to protocols that are window-based (Copa), or throttle transmissions for network RTT measurements (BBR).
    \item \emph{Competitive Throughput :} \sqp's bandwidth probing and sampling mechanism is competitive by default, and achieves high, stable throughput share when competing with queue-building flows that cause delays within an acceptable range.
    \sqp avoids high queuing delays and starving other flows using mechanisms like adaptive one-way delay measurements, a bandwidth target multiplier, and frame pacing.
    \sqp's design avoids the pitfalls of  delay-based CCAs that use explicit mode switching (e.g., Copa~\cite{arun2018copa, goyal2018elasticity-nimbus}).
    
\end{compactenum}

}

\sqp's evaluation on real-world wireless networks for \company's AR streaming platform, and across a variety of emulated scenarios, including real-world Wi-Fi and LTE traces show that:
\begin{compactenum}
    \item
    Under A/B testing of \sqp and \copa
    \footnote{Adapted from mvfst~\cite{mvfst-copa}, does not implement Copa's mode switching.}
    on \company's AR streaming platform across $\approx 8000$ individual streaming sessions, 71\% of \sqp sessions on Wi-Fi have P50 bitrate $>$ 3 Mbps and P90 frame RTT $<$ 100 ms, compared to 56\% for \copa. 
    On cellular links, 36\% of \sqp sessions meet the criteria versus only 9\% for \copa.
    \item Across emulated wireless traces, \sqp's throughput is 11\% higher than \copa (without mode switching) with comparable P90 frame delays, while \copa (with mode switching), \sprout~\cite{winstein2013stochastic}, and \bbr incur a 140-290\% increase in the end-to-end frame delay relative to \sqp. 
    \item \sqp achieves high and stable throughput when competing with buffer-filling cross traffic. Compared to \copa (with mode switching), \sqp achieves 70\% higher P10 bitrate when competing with \cubic, and 36\% higher link share when competing with \bbr.
\end{compactenum}

\textbf{This work does not raise any ethical issues.}

\section{Related Work}
\revised{
Loss-based CCAs like Reno~\cite{padhye2000modeling} and \cubic~\cite{ha2008cubic} were designed to maximize network utilization while preventing congestion collapse.
However, their buffer-filling behavior causes high queuing delay, and the bursty nature of their packet transmissions results in frames getting buffered at the sender.
These properties make these algorithms ill-suited for low-latency interactive streaming applications.
Minimizing frame latency by reducing in-network queuing, and by avoiding buffering frames at the sender, is crucial for achieving high quality of experience (QoE) for such applications, even if bandwidth is sacrificed to some extent to achieve this goal.
}
{
\newcolumntype{q}{>{\hsize=0.7\hsize}X}
\newcolumntype{w}{>{\hsize=0.55\hsize}X}
\newcolumntype{e}{>{\hsize=0.5\hsize}X}
\newcolumntype{r}{>{\hsize=0.48\hsize}X}
\newcolumntype{t}{>{\hsize=1\hsize}X}

\newcommand{\heading}[1]{\multicolumn{1}{c}{#1}}

\begin{table*}[htbp]
\centering
\resizebox{0.9\textwidth}{!}{\begin{minipage}{\textwidth}
\small{
\begin{tabularx}{\textwidth}{|q|w|e|r|t|} 
 \cline{1-5}
 \textbf{Protocol Category} &\textbf{Congestion Detection} & \textbf{Competing with Queue-building Flows} & \textbf{Congestion Control Mechanism} & \textbf{Comments} \\ \cline{1-5}
 \textbf{Explicit signaling} \newline DCTCP,  \newline ABC, XCP~\cite{alizadeh2010datadctcp,goyal2020abc,zhang2005implementationxcp} & Explicit signals from network to detect congestion & Compete with homogeneous flows & Various & Lack of support, traffic heterogeneity - unsuitable for Internet-based interactive video streaming \\
 \cline{1-5}
 \textbf{Low Delay} \newline TCP-Lola, TCP-Vegas, Sprout (Salsify), \newline TIMELY, Swift~\cite{kumar2020swift, mittal2015timely} & Packet delay/delay-gradient, stochastic throughput forecast (Sprout) & Queue-building flows cause low throughput & Window-based, Rate-based (TIMELY), hybrid (Swift) & High, stable throughput required - not achieved with queue-building cross-traffic, custom encoder for handling bursty CCA (Salsify)  \\
 \cline{1-5}
 \textbf{Mode Switching} \newline Copa, Nimbus, \newline GoogCC (WebRTC), & Packet delay/delay-gradient as congestion signal & More aggressive when competing flow detected & Window-based, Rate-based (GoogCC) & Mode-switching is unstable (Copa, GoogCC), can be slow to converge (Nimbus, GoogCC) \\
 \cline{1-5}
 \textbf{Model-Based} \newline BBR & minRTT probe, pacing gain for bandwidth & Designed to be competitive with \cubic & Rate- and window-based & 200ms minRTT probe throttles transmissions, 2 BDP in-flight under ACK aggregation/competition \\
 \cline{1-5}
 \textbf{Utility-based} \newline PCC, \newline PCC-Vivace & Explicit probing, delay, packet loss & Measure network response to rate change & Rate-based & Inconsistent performance with queue-building flows, slow convergence on dynamic links\\
 \cline{1-5}
\end{tabularx}
}
\end{minipage}}
\caption{Various CCAs that exist today, and their properties.}
\label{table:cca_background}
\end{table*}

A suitable congestion control algorithm for low-latency interactive video streaming must satisfy the four key properties discussed in \S~\ref{sec:introduction}.
Various CCAs are summarized in Table~\ref{table:cca_background}.
Low-latency CCAs like TCP-Lola~\cite{hock2017tcp-lola}, TCP-Vegas~\cite{brakmo1994tcp-vegas}, and Sprout (Salsify\footnote{Salsify streamer uses Sprout as the CCA (used interchangeably)})~\cite{winstein2013stochastic, fouladi2018salsify} that use packet delay as a signal have a key drawback: they are unable to achieve high throughput when competing with queue-building cross-traffic.
Some mode switching algorithms (e.g., Copa~\cite{arun2018copa}) can misinterpret self-induced queues as competing flows, resulting in high delays, whereas other CCAs like Nimbus~\cite{goyal2018elasticity-nimbus}, and GoogCC (WebRTC)~\cite{carlucci2016analysis, webrtc-eval} converge slowly, and can have unstable throughput when competing with queue-building flows.

BBR~\cite{cardwell2016bbr} periodically throttles transmissions to measure a baseline delay, which is problematic for interactive video streaming since frames cannot be transmitted during its minRTT probe.
Window-based protocols have a similar problem - they transmit packets in bursts, and a mismatch between packet transmissions and frame generation require sender-side buffering, and increase the end-to-end frame delay.

Utility-based algorithms like PCC~\cite{dong2015pcc, dong2018pcc} explicitly probe the network and aim to maximize a utility function based on throughput, delay, and packet loss.
These CCAs converge slowly on dynamic links, and have inconsistent performance when competing with queue-building flows.

CCAs use rate-based or window-based mechanisms in order to control the transmission rate under congestion. 
Rate-based CCAs are better suited for video streaming due to the burst-free nature of packet transmissions, whereas the bursty window-based mechanisms can block frame transmissions at the sender and make encoder rate-control challenging (e.g., Salsify-Sprout). The other benefit of rate-based CCAs is that their internal bandwidth estimate can be used to directly set the video bitrate, whereas window-based mechanisms require additional mechanisms for setting the video bitrate. \srini{ends a bit abruptly}\dd{PTAL, not happy with it, any suggestions?}



}

\revised{
In recent years, there has been an increased interest in congestion control algorithms that reduce self-induced in-network queuing.
Delay-based algorithms like \copa (without mode switching)~\cite{arun2018copa}, TCP-Vegas~\cite{brakmo1994tcp-vegas}, TCP-LoLa~\cite{hock2017tcp-lola}, and TCP-FAST~\cite{wei2006fast} are promising advances in this space. \sprout~\cite{winstein2013stochastic} achieves high throughput and low delay on noisy links with delay jitter (including widely deployed Wi-Fi, cellular, and DOCSIS technologies) using stochastic bandwidth forecasts.
Unfortunately, their throughput is low in the presence of queue-building cross traffic like \cubic and \bbr~\cite{cardwell2016bbr}, and are thus not suitable for interactive streaming.
}
{
}

\revised{
To overcome some of the limitations of low-delay congestion control algorithms, \copa~\cite{arun2018copa} implements a mode switching mechanism that makes \copa more aggressive when it detects buffer-filling cross traffic.
Any algorithm with explicit mode switching must choose a threshold for switching between modes - this inherently makes these designs fragile since misdetection can result in high delays in the absence of buffer-filling cross traffic or low throughput in the presence of buffer-filling cross traffic.
}
{


}

\revised{
\bbr~\cite{cardwell2016bbr} is a hybrid rate-delay model-based algorithm designed to achieve high throughput and lower delay compared to loss-based CCAs, but its inclusion of a 200 ms PROBE\_RTT phase for periodically measuring an RTT baseline would pause the transmission of frames at the sender, making \bbr unsuitable for low-latency interactive streaming applications (\S~\ref{sec:prelim_study}).
In addition, overestimation of the available bandwidth can cause additional queuing, resulting in up to 1 RTT of extra queuing delay (\bbr may maintain up to 2 BDP in-flight~\cite{ware-bbr}, shown in Figure~\ref{fig:background:delay_timeseries}).

Another category of congestion control algorithms that has been recently proposed is based on maximizing some notion of overall utility in the presence of congestion.
\pcc~\cite{dong2015pcc} and \pcc-Vivace~\cite{dong2018pcc} run online experiments by changing the send rate and measuring the utility, aiming to converge to a behavior that maximizes network utility.
In our experiments (\S~\ref{sec:evaluation}), we observe that these algorithms are slow to converge to the correct rate.
}
{
}
\section{Preliminary Study}
To illustrate the shortcomings of existing congestion control algorithms in the context of low-latency interactive streaming, we present some preliminary experimental results using the Pantheon~\cite{yan2018pantheon, netravali2015mahimahi} testbed.
Details regarding the specific CCA implementations are provided in \S~\ref{sec:eval:setup}.

\subsection{Variable Bandwidth Link}
\label{sec:prelim_study}

\begin{figure}
    \centering
    \includegraphics[width=\columnwidth]{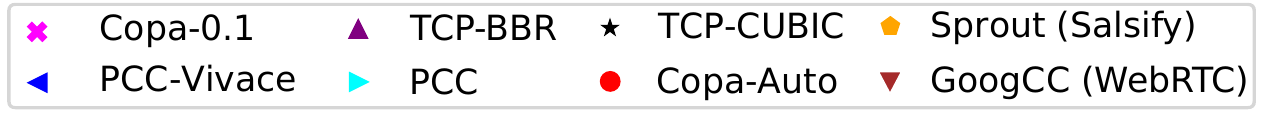}
    \subcaptionbox{Send rate timeseries.\label{fig:background:converged_timeseries}}{
    \includegraphics[width=0.47\columnwidth]{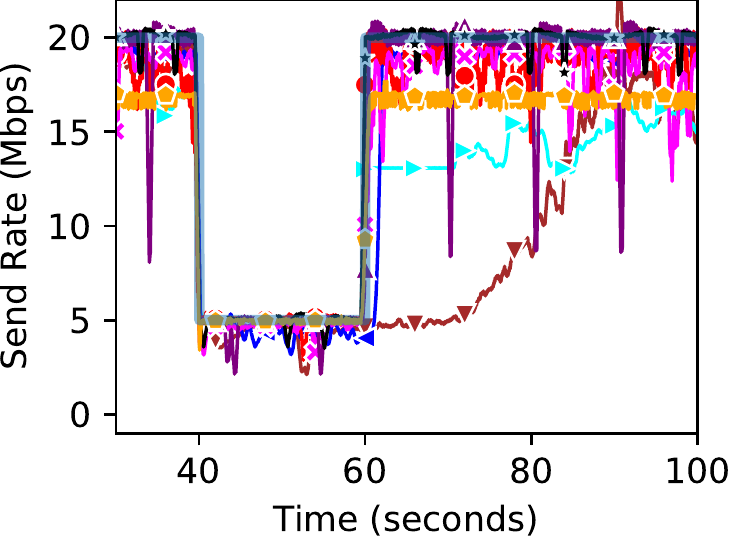}
    }%
    \hfill
    \subcaptionbox{Packet delay timeseries.\label{fig:background:delay_timeseries}}{
    \includegraphics[width=0.47\columnwidth]{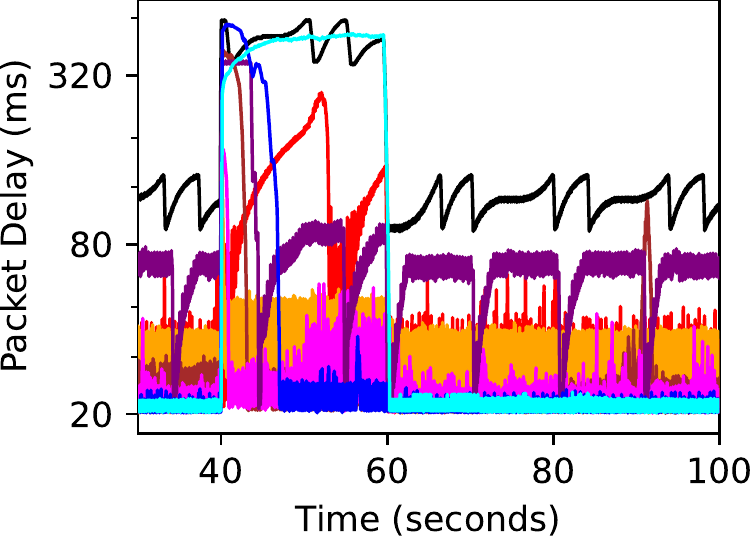}
    }%
    \caption{Congestion control performance on a variable link (link bandwidth shown as a shaded light blue line).}
    \label{fig:background:experiment_full}
\end{figure}
\revised{
We ran a single flow for 120 seconds over a bottleneck link that periodically oscillates between \SI{15}{Mbps} and \SI{20}{Mbps}, and has a baseline RTT of 40 ms.
}
{
We ran a single flow for 120 seconds over a 40 ms RTT link, where the bandwidth temporarily drops down to \SI{5}{Mbps} from \SI{20}{Mbps}.
}
The goal of this experiment is to see if the algorithm can (1) quickly discover additional bandwidth when the available bandwidth increases, and (2) maintain low delay when the available bandwidth decreases.

\revised{
Figure~\ref{fig:background:experiment_full} shows the throughput and delay timeseries for the last 20 second period for each algorithm.
}{
Figure~\ref{fig:background:experiment_full} shows the throughput and delay timeseries between $t=30s$ and $t=100s$.
}
\revised{
While both \copa-0.1 (\copa without mode switching with $\delta = 0.1$ \footnote{A lower delta makes \copa more aggressive, sacrificing low delay for higher throughput. The original paper proposes using 0.5, whereas the version on Pantheon uses 0.1. Facebook's testing of \copa~\cite{garg-2020} used 0.04. }) 
and \copa-Auto (Copa with mode-switching using adaptive $\delta$) can reasonably track the link bandwidth, they are susceptible to occasional drops in the throughput.
In addition, both \copa-0.1 and \copa-Auto have delay variation due to \copa's 5-RTT probing cycle, which temporarily injects network queuing and subsequently drains it to probe for additional bandwidth, where the delay is inversely proportional to the value of $\delta$.
The delay variation is worse in the case of \copa-Auto since it periodically misdetects the presence of buffer-filling cross traffic and reduces the value of $\delta$ to prioritize bandwidth over delay.
\sprout is able to utilize the link when its capacity is \SI{15}{Mbps}, but fails to utilize the link when the link capacity goes up to \SI{20}{Mbps}.
In addition, Sprout has a higher peak queuing delay compared to \copa-0.1.

\pcc, \vivace and \webrtc react slowly to changes in the link bandwidth, resulting in severe underutilization of the link (Figure~\ref{fig:background:converged_timeseries}) and large delay spikes (Figure~\ref{fig:background:delay_timeseries}) respectively.

While \bbr and \cubic utilize the link completely, they suffer from significantly higher queuing delays due to queue-building behavior (Figure~\ref{fig:background:delay_timeseries}).
While this is expected behavior since both \bbr and \cubic prioritize throughput over low delay, this makes the algorithms not suitable for low-latency interactive streaming applications.

}{
Throughout the entire trace, \cubic operates with the queues completely full, resulting in high link utilization at the cost of high queuing delays.
Among the low-delay algorithms, the delay performance differs greatly across specific algorithms.
When the link rate goes down to \SI{5}{Mbps}, \sprout and \copa-0.1 (\copa without mode switching, $\delta=0.1$\footnote{A lower delta makes \copa more aggressive, sacrificing low delay for higher throughput. The original paper proposes using 0.5, whereas the version on Pantheon uses 0.1. Facebook's testing of \copa~\cite{garg-2020} used 0.04.}) are able to adapt rapidly without causing a delay spike.
\pcc-Vivace~\cite{dong2018pcc}, GoogCC and \bbr are slower to adapt, causing 3-8 seconds long delay spike.
Throughout the low bandwidth period, \pcc maintains persistent, high queuing delay, whereas \copa-Auto (Copa with mode-switching, adaptive $\delta$) incorrectly switches to competitive mode, significantly increasing queuing delay.

In addition to the delay that occurs when the link rate drops, some algorithms have inherently more queuing than others.
\bbr can maintain up to 2 BDP in-flight, causing up to 1 BDP of in-network queuing.
Both, \copa-0.1 and \copa-Auto demonstrate significant short-term delay variations due to \copa's 5-RTT probing cycle, which serves the role of probing the network for additional capacity.
The peak delay is inversely proportional to the value of $\delta$, and is worse in the case of \copa-Auto, since it periodically misinterprets its own delay as delay caused due to a competing queue-building flow, and consequently reduces the value of $\delta$ in response.
There is significant variation in \sprout's packet delay due to the bursty nature of its packet transmissions, even though it is significantly underutilizing the link.
GoogCC also demonstrates a delay spike around $t=90s$, when its send rate hits the link limit after an extended ramp-up period.
\pcc-Vivace, \copa-Auto, \copa-0.1, and \bbr are able to rapidly probe for more bandwidth when the link rate increases.
On the other hand, \pcc and GoogCC are the slowest to converge, taking more than 20-30 seconds to ramp up after the link rate increases, resulting in severe underutilization.
}

\revised{
\sqp measures network bandwidth by pacing packets in sync with the video frames (\S~\ref{sec:pacing_target} and always paces packets faster than the video bitrate. Avoiding unnecessary changes in the pacing rate (unlike \bbr, \copa, \pcc, and \vivace) helps avoid queuing delays at the sender.  
}
{
In order to achieve high link utilization and low delays for low-latency interactive video streaming, the CCA must quickly identify the link capacity without causing queuing delays, and quickly back off when the delay is self-induced. 
\sqp is able to achieve these requirements, as shown in Section~\ref{sec:appendix:timeseries}.
}

\subsection{Short Timescale Variations}
\label{sec:background:short_var}
\begin{figure} 
    \centering
    \subcaptionbox{Short-term delay variation.\label{fig:background:delay_zoomed}}{
    \includegraphics[width=0.47\columnwidth]{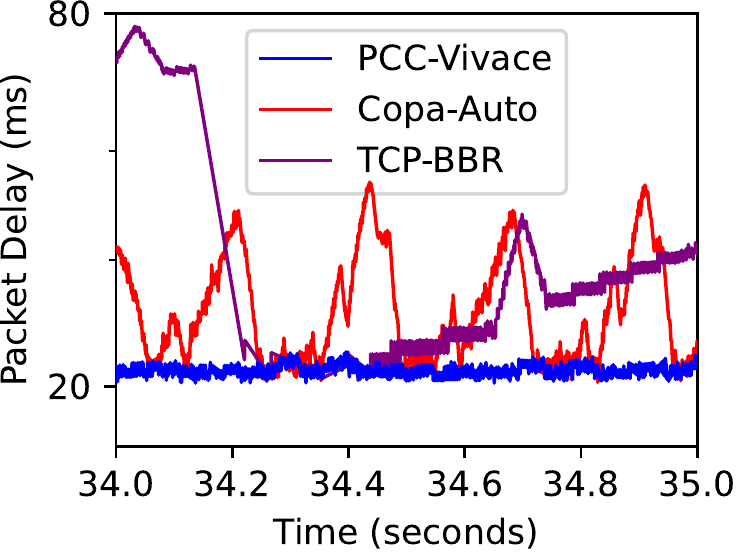}
    }%
    \hfill
    \subcaptionbox{Short-term send rate variation.\label{fig:background:send_var}}{
    \includegraphics[width=0.47\columnwidth]{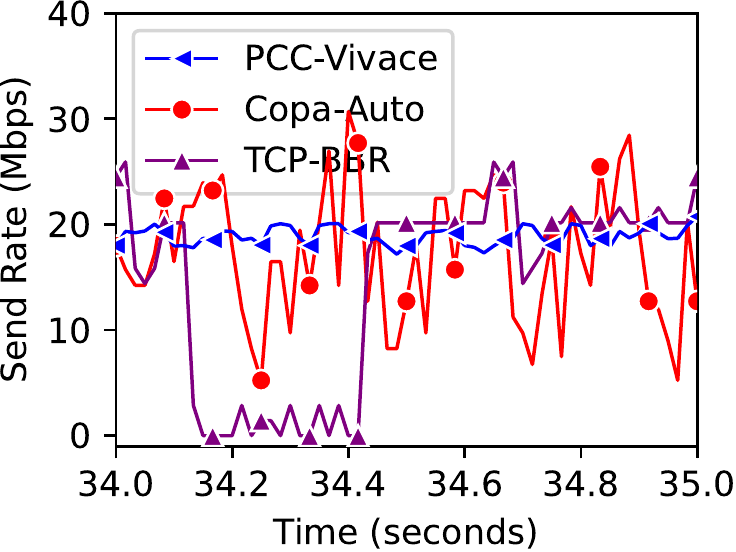}
    }
    \caption{A closer look at the short term delay and send rate variation on a constant 20 Mbps link.}
    \label{fig:background:experiment}
\end{figure}

\revised{
The complex interactions between the congestion control layer and the encoder play a major role in the end-to-end performance of the streaming application. 
}{}
In this section, we examine the short timescale behavior of existing protocols to see if they can provide the low packet delay and stable throughput~\cite{tfrc} needed to support the requirements of low-latency streaming applications.
We present three algorithms that demonstrate distinct short-term behavior: \copa-Auto, \bbr and Vivace (additional results in Section~\ref{sec:appendix:shortvar}).
\copa-0.1 and \sprout behave similar to \copa-Auto, and \pcc behaves similar to Vivace in these experiments.
We ran each algorithm on a fixed 20 Mbps link with 20 ms of delay in each direction.
Figure~\ref{fig:background:delay_zoomed} shows the one-way packet delays and Figure~\ref{fig:background:send_var} shows the packet transmission rate for each frame period (16.66 ms at 60FPS).

\revised{
\copa-Auto's one-way packet delay oscillates between 20 ms and 60 ms over a 12-frame period. Its send rate also varies significantly at the frame level.
}{
\copa-Auto's one-way packet delay oscillates between 20 ms and 60 ms over a 12-frame period, with large variations in the send rate at frame-level timescales.
}
If a smooth video bitrate is determined using the average send rate to maximize utilization, the frames at $T=34.3,~34.5$ would get delayed at the sender. 
To lower the sender-side queuing delay, the encoder rate selection mechanism must either: (1) choose a conservative video bitrate, resulting in underutilization, or (2) have frequent rate control updates.

While \bbr is not particularly suitable for interactive streaming because of its higher queuing delay, \bbr's RTT probing mechanism is especially problematic.
Every 10 seconds, \bbr throttles transmissions (transmitting at most 4 packets per round trip) for 200 ms to measure changes in the link RTT (between $T=34.2$ and $T=34.4$).
During this period the generated video frames will be queued at the sender, resulting in 200 ms of video stutter every 10 seconds.

\revised{Algorithms like \pcc and \vivace directly model a send rate based on network observations, and rate-limit packet transmissions using pacing.
While such designs are better suited for streaming applications since the algorithm's internal rate tracking mechanism can be used to set the encoder bitrate, factors like encoder overshoot can still cause sender side queuing delays.
In addition, the network state may change in the few milliseconds it takes to encode a frame, causing the congestion control algorithm to delay transmitting the frame.
}
{
Rate-based algorithms like \pcc and \vivace are better suited for streaming applications, since the internal rate-tracking mechanism can be used to set the video bitrate, and frames are not delayed at the sender if the encoded frames do not overshoot the requested target bitrate.
While Salsify~\cite{fouladi2018salsify} attempts to solve this problem using a custom encoder that can match the instantaneous transmission rate of a bursty CCA like Sprout~\cite{winstein2013stochastic}, rearchitecting the CCA is a more universal solution that can leverage advances in hardware video codecs that have good rate control mechanisms  (App.~\ref{sec:appendix:encoderovershootundershoot}).
}

\revised{
\sqp directly measures network bandwidth by pacing packets in sync with the video frames  (\S~\ref{sec:pacing_target}) and always paces packets faster than the video bitrate.
Avoiding unnecessary changes in the pacing rate (unlike \bbr, \copa, \pcc, and \vivace) helps avoid queuing delays at the sender.  
}
{
To minimize the end-to-end frame delay, the CCA must transmit encoded frames immediately, and pace faster than the rate at which the network can deliver the packets.
\sqp directly controls the video bitrate using smooth bandwidth estimates, and the transmissions are synchronized with the frames, which reduces the end-to-end frame delay (Section~\ref{sec:appendix:shortvar}).
}



\section{Design}
\label{sec:design}


\revised{
A typical low-latency interactive streaming application generates raw frames at a fixed frame-rate according to inputs received from the client device.
These frames are compressed by the encoder according to a bitrate chosen by an adaptive bitrate (ABR) algorithm that works in conjunction with the transport layer congestion control algorithm to manage the end-to-end frame delay, encoder bitrate churn and the bandwidth utilization.
Compressed frames are then queued for transmission and transmitted by the transport layer in a manner that attempts to avoid network congestion, and are eventually decoded and displayed at the client device.
}
{
Low-latency interactive streaming applications generate raw frames at a fixed frame-rate.
The video bitrate is determined by an adaptive bitrate (ABR) algorithm using signals from the CCA in order to manage frame delay, network congestion, and bandwidth utilization.
The compressed frames are transmitted over the network, and eventually decoded and displayed at the client device.
}

\revised{
In this section, we describe the design of \sqp, a rate- and delay-based congestion control algorithm for low-latency real-time video streaming that tightly couples video content transmission and network congestion control.
Integration with the video transmission reduces sender-side frame transmission delays by closely matching the congestion control algorithm's transmission rate and the video bitrate.
\sqp's bandwidth estimation mechanism is based on frame pacing and piggy-backs network measurements onto video frame transmissions, achieving a competitive throughput share when sharing a bottleneck with queue-building cross-traffic, while simultaneously achieving low end-to-end frame delay in isolation.
}{
\sqp is a rate- and delay-based CCA for low-latency interactive video streaming, and aims to (1) provide real-time bandwidth estimates that ensure high utilization and low end-to-end frame delay on highly variable links, and (2) achieve competitive throughput in the presence of queue-building cross traffic.
\sqp's congestion control mechanism must be purely rate-based in order to avoid the undesirable trade-offs between bandwidth utilization, encoder bitrate changes, and the end-to-end frame delay (\S~\ref{sec:introduction}).

}

\subsection{Architecture Overview}
\label{sec:overview}

\begin{figure}
    \centering
    \subcaptionbox{Video streaming architecture.
    \label{fig:design:overall_architecture}}{
    \includegraphics[width=0.9\columnwidth]{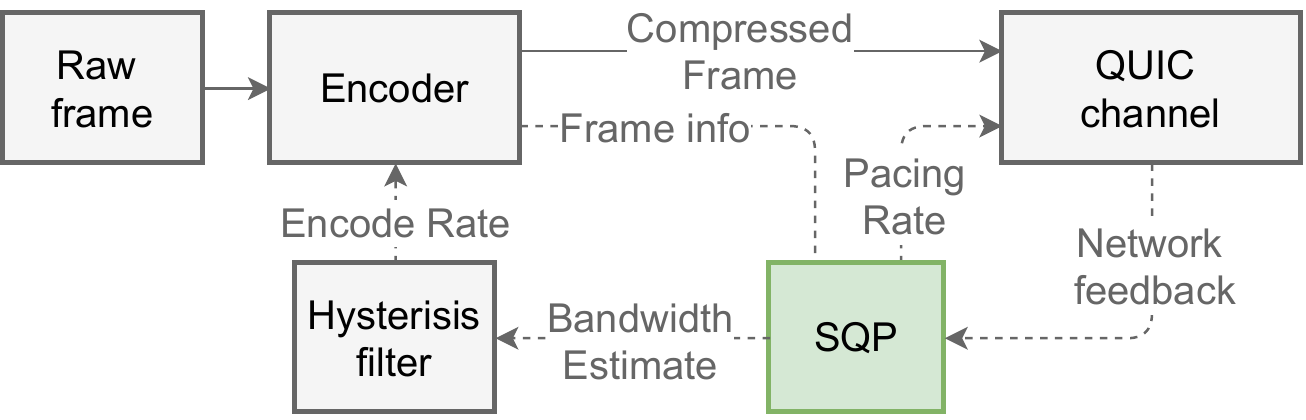}
    }%
    \hfill
    \subcaptionbox{\sqp internal architecture. \label{fig:design:sqp_architecture}}{
    \includegraphics[width=0.9\columnwidth]{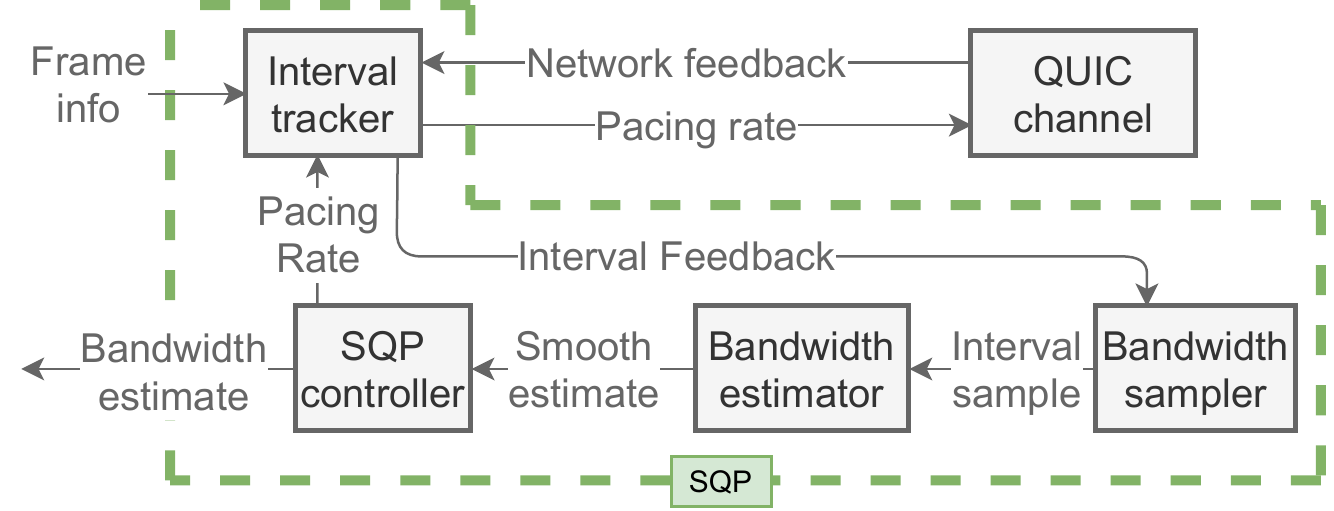}
    }%
    \caption{Low-latency video streaming and \sqp architectures. 
    }
\end{figure}

\sqp's role in the end-to-end streaming architecture and its key components are shown in Figures~\ref{fig:design:overall_architecture} and~\ref{fig:design:sqp_architecture}.
\sqp relies on QUIC~\cite{langley2017quic} to reliably transmit video frames, perform frame pacing, and provide  packet timestamps for estimating the network bandwidth.
\sqp directly controls the video bitrate, and a simple hysteresis filter serves as a bridge between \sqp and the encoder to reduce the frequency of bitrate changes.

\revised{
Internally, \sqp comprises of the following components:

\begin{compactenum}
\item {\em Interval tracker: } \sqp's interval tracking mechanism tracks frames pending delivery and the associated packet transmissions, to allow computing the delay and bandwidth for each frame's delivery over the network.
\item {\em Bandwidth sampler: } \sqp uses frame delivery statistics from the interval tracker to sample the available network bandwidth (\S~\ref{sec:bw_sampling}).
\item {\em Bandwidth estimator: } \sqp's bandwidth estimation algorithm computes a smooth bandwidth estimate for setting the video bitrate. It is based on network utility maximization~\cite{kelly1998rate}, ensuring fast convergence to fairness (\S~\ref{sec:bw_interp}).
\item {\em SQP Controller: } SQP's controller computes a pacing rate for each frame that is a small multiple of the bandwidth estimate. \sqp's frame pacing mechanism briefly builds short queues and quickly drains this queue during the transmission gaps between frames, enabling \sqp to implicitly probe the available network bandwidth. This is key to achieving high throughput share and friendliness when competing with queue-building cross traffic while maintaining low end-to-end frame delay (\S~\ref{sec:pacing_target}).
\end{compactenum}
}
{
Internally, \sqp's components work together in order to achieve the key design goals:
\begin{compactenum}
\item {\em Bandwidth Probing: } 
\sqp transmits each frame as a short, paced burst, and the bandwidth sampler uses frame-level packet dispersion statistics from the interval tracker for discovering additional bandwidth.
\srini{this point is not as clear as the below -- should it be a separte bullet or part of the para before -- it seems like the next few items are all about BW probing}\dd{PTAL}
\item {\em Recovery from Transient Queues: } \sqp's bandwidth samples are penalized when the delay increases over a short period (\S~\ref{sec:bw_sampling}), enabling it to recover from transient self-induced queuing.
\item {\em Recovery from Standing Queues: } \sqp uses a target multiplier mechanism (\S~\ref{sec:pacing_target}) to maintain some slack in the link utilization, enabling recovery from self-induced standing queues. 
\sqp remains competitive when competing flows cause standing queues (within acceptable delay limits) since it uses a small, dynamic window to track the transient delay (\S~\ref{sec:min_owd_window}).
\item {\em Rate-based congestion control: } \sqp aims to carefully pace each frame faster than the rate at which the network can deliver the packets (\S~\ref{sec:pacing_target}), and responds to congestion by smoothly changing the bandwidth estimate (and consequently, the video bitrate) using gradient-based updates (\S~\ref{sec:bw_interp}).
As opposed to using ACK-clocking and window-based mechanisms, rate-based congestion control simplifies integration with the video encoder and reduces the end-to-end frame delay (\S~\ref{sec:background:short_var}).
\item {\em Fairness and Interoperability: } \sqp's bandwidth estimator (\S~\ref{sec:bw_interp}) is based on maximizing a logarithmic utility function, which improves dynamic fairness due to its AIMD-style updates (\S~\ref{sec:intraprotocol_fairness}).
\sqp's frame pacing and the bandwidth target multiplier mechanisms ensure dynamic fairness across multiple \sqp flows (\S~\ref{sec:intraprotocol_fairness}), and provide a theoretical upper bound on \sqp's share when competing with elastic flows (\S~\ref{sec:bw_samp_analysis}). 
\srini{should this come last to reflect order}\dd{PTAL}

\end{compactenum}

}

\revised{
Changing the video encoding bitrate is the only control action taken by \sqp in response to congestion.
Once a frame has been encoded, \sqp paces the packets of the frame at a rate such that the packets for each frame are transmitted well before the next frame is encoded, and aims to be faster than the rate at which the network itself can deliver the packets.
This ensures that encoded frames are immediately sent onto the network, thus reducing the end-to-end frame delay by eliminating sender-side buffering. 
The network can often tolerate short-lived spikes in network traffic as long as there is sufficient bandwidth, or there is sufficient buffer space at the bottleneck and the sender subsequently takes corrective action to drain the excess queue.
\sqp takes corrective action by reducing the encoder bitrate when persistent self-induced congestion is indicated by the bandwidth samples.
Thus, the transmission rate for packets is limited by the generated video bitrate, which is in contrast to algorithms that use ACK-clocking, congestion windows, and pacing-based rate limiting.
The frame pacing mechanism is discussed in \S~\ref{sec:pacing_target}.
}
{

}

For the initial part of the discussion, we will assume the existence of a `perfect' encoder with the following properties: (1) the target bitrate can be changed on a per-frame basis without any negative consequences, as long as the target bitrate does not change significantly from frame to frame, and (2) the encoder does not overshoot or undershoot the specified target rate.
In \S~\ref{sec:encoder_variation}, we discuss how \sqp works in a practical setting when encoders do not satisfy some of these assumptions.

\subsection{Bandwidth Sampling}
\label{sec:bw_sampling}
\begin{figure*}
    \centering
    \subcaptionbox{\sqp's bandwidth samples are higher than the video bitrate when the link is underutilized, indicating that the video bitrate should be increased.
    \label{fig:design:underutilization}}{
    \includegraphics[width=0.3\textwidth]{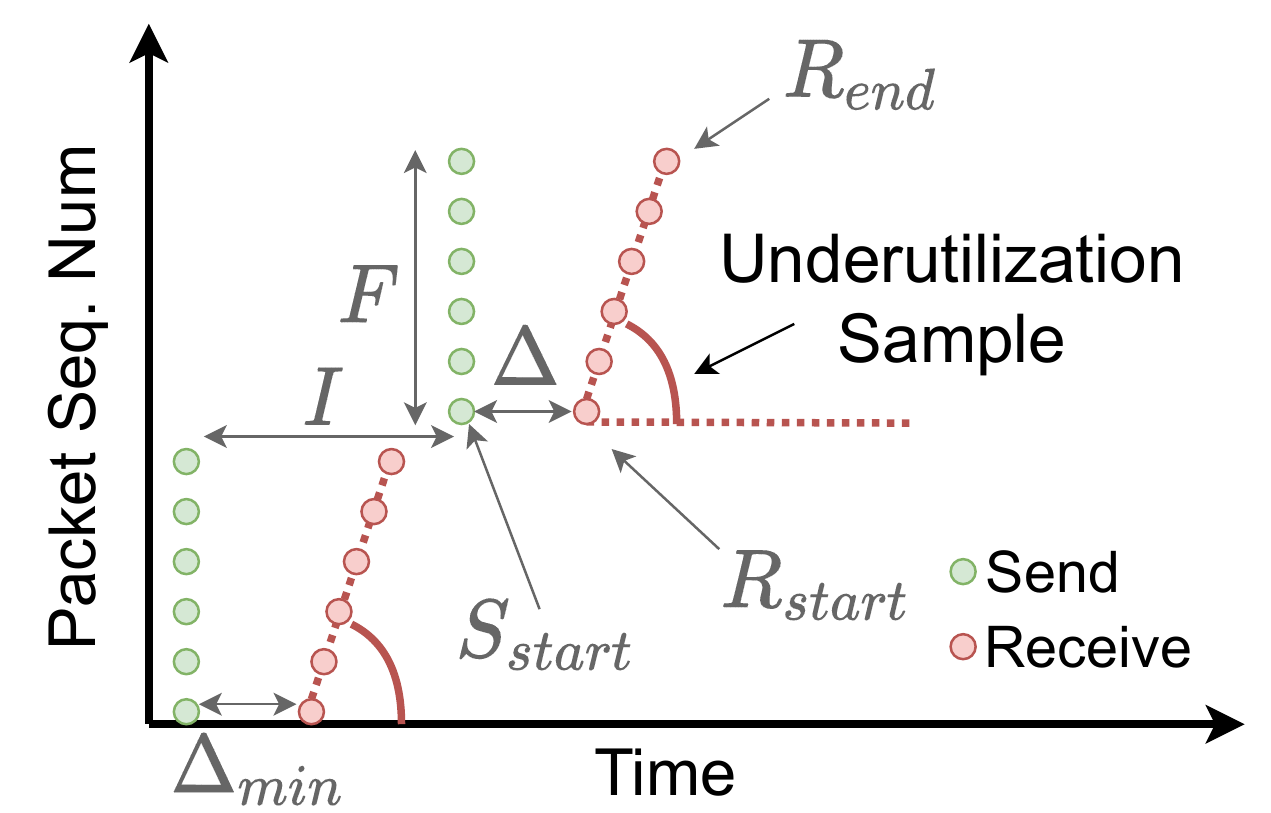}
    }%
    \hfill
    \subcaptionbox{\sqp's bandwidth samples are lower than the link rate when the link is overutilized, indicating that the video bitrate should be reduced.\label{fig:design:overutilization}}{
    \includegraphics[width=0.3\textwidth]{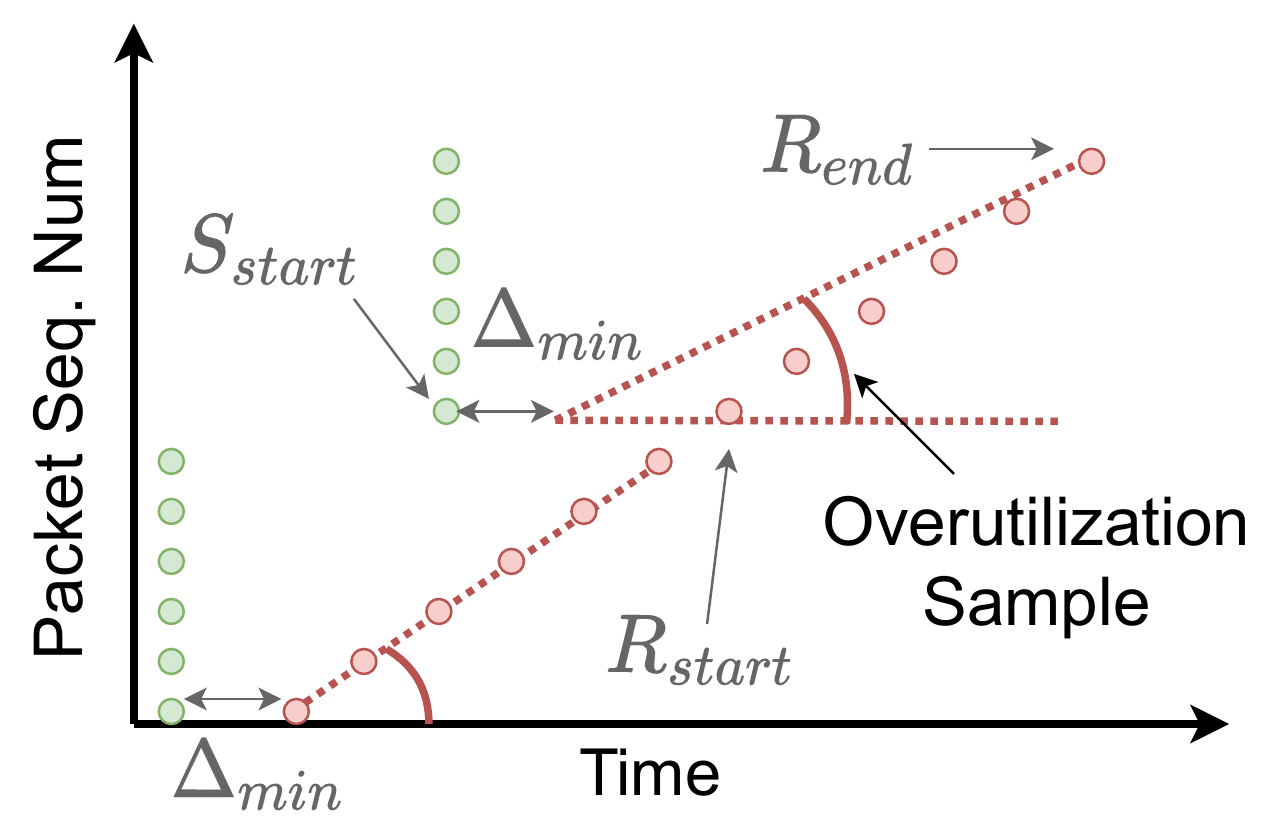}
    }%
    \hfill
    \subcaptionbox{Bandwidth samples from frames that are smaller than the target bitrate
    \srini{small frames or smaller than expected} \dd{PTAL}
    are more sensitive to transient queuing. \sqp's corrected bandwidth sample is closer to the link rate than the uncorrected sample.\label{fig:design:app_limited}}{
    \includegraphics[width=0.3\textwidth]{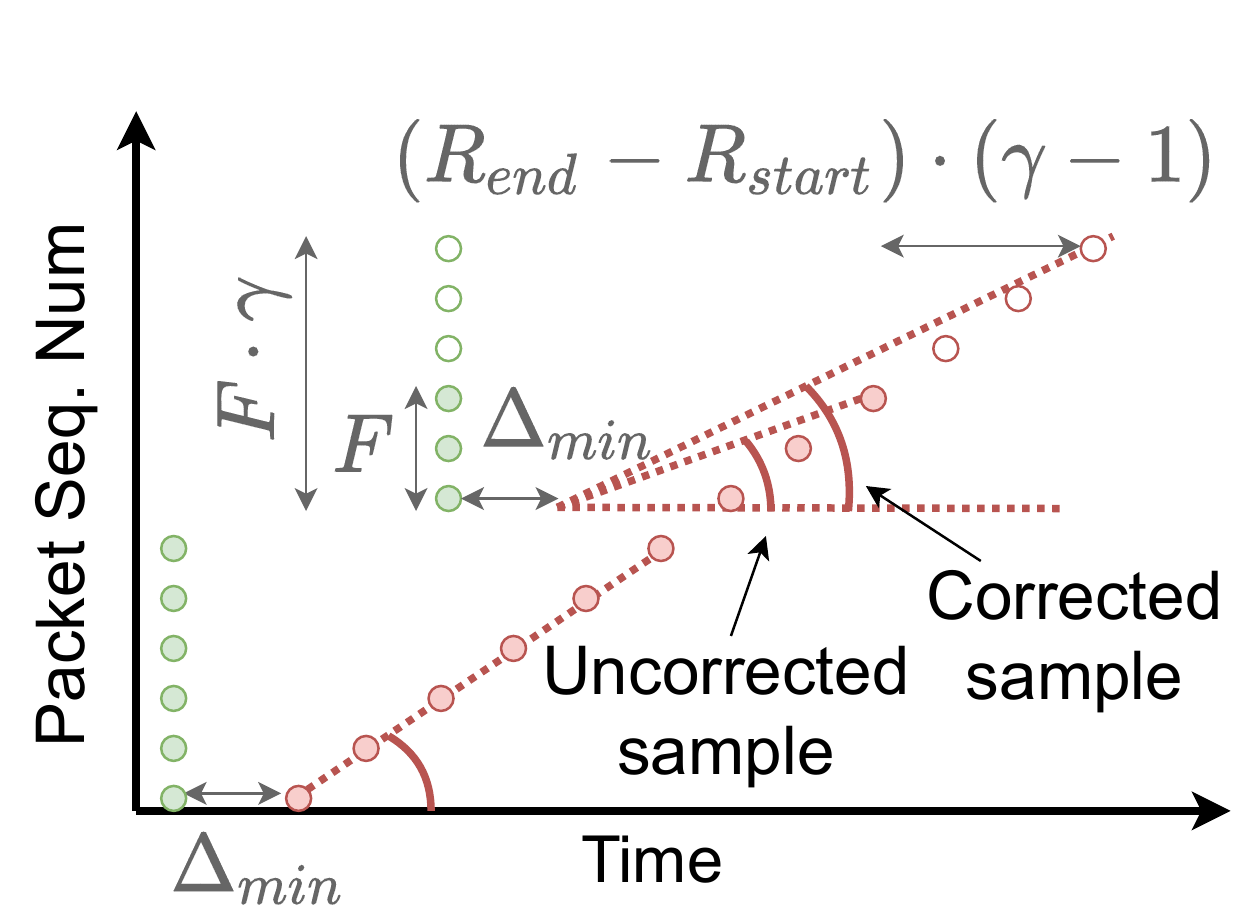}}%
    
    \caption{\sqp's bandwidth samples converge towards the link rate and aid in draining self-inflicted queues. The slope of the dotted red line represents the bandwidth sample in each case.}
    \label{fig:design:bw_sampling}
\end{figure*}

The goal of \sqp's bandwidth sampling algorithm is to measure the end-to-end frame transport rate that achieves high link utilization while avoiding self-induced queuing and packet transmission pauses (e.g., \copa and \bbr in \S~\ref{sec:background:short_var}).
\sqp transmits each frame as a short burst that is faster than the network delivery rate, which causes a small amount of queuing. This queue is drained by the time the next frame arrives at the bottleneck if the average video bitrate is lower than the available bottleneck link capacity.
\sqp uses the dispersion of the frame-based packet train~\cite{dovrolis2004packet} to measure link capacity, with some key differences that aid in congestion control compared to basic packet-train techniques.
\sqp's probing works at sub-frame timescales ($<16.66~\mathrm{ms}$ @ 60FPS), in contrast to CCAs that probe for bandwidth over longer timescales (\pcc:2RTT, \bbr:1 min-RTT, and \copa:2.5 RTT).

Consider an application generating frames at a fixed frame rate.
As shown in Figure~\ref{fig:design:bw_sampling}, a frame of size $F$ is transmitted every inter-frame time interval, $I$ (e.g., 16.66 ms at 60FPS), and the average bitrate is $\frac{F}{I}$.
Each frame is paced at a rate that is higher than $\frac{F}{I}$ (shown by the steep slope of the green dots).
If there are no competing flows (competition scenario discussed in \S~\ref{sec:bw_samp_analysis}), and the link bandwidth is lower than the pacing rate, the packets will get spaced out according to the bandwidth of the bottleneck link (slope of the red dots). \sqp computes the end-to-end frame transport bandwidth sample as: 
\srini{should add a sentence or two clarifying spaced by capacity or available BW -- and how pacing rate may impact this}\dd{PTAL}
\begin{align}
    \label{eq:bw_samp_owd}
    S = \frac{F}{R_{\mathit{end}} - S_{\mathit{start}} - \Delta_{\mathit{min}}}
\end{align}
This is the slope of the red dotted line in Figure~\ref{fig:design:bw_sampling}. 
$S_{\mathit{start}}$ and $R_{\mathit{end}}$ are the send and receive times of the first and last packets of a frame, respectively, and $\Delta_{\mathit{min}}$ is the minimum one-way delay (delta between send and receive timestamps) for packets sent during a small window in the past (\S~\ref{sec:min_owd_window}).
$\Delta_{\mathit{min}}$ and $R_{\mathit{end}} - S_{\mathit{start}}$ have the same clock synchronization error (sender-side vs. receiver-side timestamps) and cancel each other out. 
\srini{I think give forward pointers for both pacing gain and tracking window here}\dd{PTAL}

\textbf{Underutilization Sample:} 
When the network is underutilized or 100\% utilized, no additional queuing occurs across multiple frames ($\Delta = R_{start}-S_{start} = \Delta_{\mathit{min}}$ remains constant). 
Thus, as shown in Figure~\ref{fig:design:underutilization}, the sample is equal to the packet receive rate for a frame (ie. the bottleneck link bandwidth).
The samples during link underutilization are higher than the video bitrate ($\frac{F}{I}$), and \sqp increases its bandwidth estimate.
When the link is 100\% utilized, the samples are equal to the video bitrate, indicating good link utilization.

\textbf{Overutilization Sample:} 
Transient overutilization due to frame size overshoots, bandwidth overestimation (link aggregation, token bucket policing), or a drop in network bandwidth can cause queuing that builds up across multiple frames.
This results in an increase in $\Delta - \Delta_{min}$ for subsequent frames, which lowers the bandwidth samples for subsequent frames (Figure~\ref{fig:design:overutilization}, slope of dotted red line for the second frame). 
Thus, \sqp lowers the video bitrate below the link rate and recovers from transient queuing.
\revised{}{
When packets are lost, \sqp scales down its samples by the fraction of lost packets, primarily responding to sustained loss events (e.g., shallow buffers, \S~\ref{sec:eval:shallow_buf}).
}

\revised{}{
\srini{why does delta min need to be low -- the tracking window discussion is a bit unclear at this point} \dd{Would it be better to merge this with 4.3?} \neal{yes i'd vote to merge this into 4.3}\dd{Done}
}

\revised{
}{
\textbf{Video Encoder Undershoot:}
While \sqp is also able to discover the link bandwidth quickly in application-limited scenarios since it relies on the pacing burst rate, and not the average video bitrate, bandwidth samples from small frames are unfairly penalized due to delay variations.
\sqp often has to deal with application-limited scenarios where the bitrate of the encoded video is less than the bandwidth estimate.
This can be due to conservative rate control mechanisms that serve as a bridge between the bandwidth estimate and the encoder bitrate to reduce the frequency of encoder bitrate updates, or due to encoder undershoot during low complexity scenes that do not warrant encoding frames at the full requested target bitrate (eg. low-motion scenes like menus).
When \sqp is application-limited, queuing delay from past frames can unfairly penalize the bandwidth sample (Figure~\ref{fig:design:app_limited}). \srini{figure looks like a frame that takes more than inter-frame gap to transmit -- why is this small?}\dd{The first frame is a large one that caused transient queuing, the second one is a smaller frame.}
While padding bytes can be used to bring up the video bitrate to \sqp's bandwidth estimate, this results in wastage of bandwidth.
To improve \sqp's robustness under application-limited scenarios, we modify the bandwidth sampling equation to account for undershoot:
\begin{align}
    \label{eq:bw_samp_undershoot}
    S = \frac{F \cdot \gamma}{R_{\mathit{end}} - S_{\mathit{start}} - \Delta_{\mathit{min}} + (R_{end}-R_{start})\cdot (\gamma -1)}
\end{align}
where $\gamma = \frac{F_{max}}{F}$ is the undershoot correction factor, $F_{max}$ is the hypothetical frame size without undershoot, and $(R_{end}-R_{start})\cdot (\gamma -1)$ is the predicted additional time required for delivering the hypothetical full-sized frame.
This computes the bandwidth sample by extrapolating the delivery of a small frame to the full frame size that is derived from \sqp's current estimate. In Figure~\ref{fig:design:app_limited}, the solid dots are actual packets for a frame, and the hollow dots show the extrapolated transmission and delivery of the packets.
}

\subsection{Tracking Minimum One-way Delay ($\Delta_{min}$)}
\label{sec:min_owd_window}
\revised{
The window size for tracking $\Delta_{min}$ represents the duration of \sqp's memory of the minimum packet transmission delay.
$\Delta_{min}$ serves as a baseline for detecting self-inflicted network queuing, and the window size affects \sqp's throughput when competing with queue-building cross traffic.
}
{
}
\revised{

}{
}

\revised{


The minimum packet transmission delay, $\Delta_{min}$, serves as a baseline for detecting self-inflicted network queuing. 
The window size for tracking $\Delta_{min}$ represents the duration of \sqp's memory of $\Delta_{min}$, which affects \sqp's throughput when competing with queue-building cross traffic.

If a very small window is used (e.g., 0.1-0.5~s), $\Delta_{\mathit{min}}$ quickly increases in response to an increase in standing queue.
Since \sqp injects packets into the network at a rate faster than \sqp's current share, it is able to probe for more bandwidth when competing with queue-building flows, even if the combined link utilization of \sqp and the cross traffic is close to 100\%. 
On the other hand, a small window implies that when self-induced queuing occurs, $\Delta_{\mathit{min}}$ can expire before \sqp can recover.

While a larger window (e.g., 10-30~s) would aid recovery from self-induced queues by anchoring \sqp to the lowest one-way delay observed over the window, \sqp's bandwidth samples would be more sensitive to delay variations caused by the queue-building cross traffic, lowering the throughput share.

\sqp uses an adaptive window size of $2 \times \mathrm{sRTT}$~\cite{paxson2000computing}, which enables \sqp to obtain a high throughput share when competing with queue-building flows that (1) do not cause very high delays, and (2) have low queuing delay variation over periods of $2 \times \mathrm{sRTT}$ (\S~\ref{sec:eval:cross_traffic}).
When self-induced queuing occurs, \sqp's adaptive window grows quickly, and in conjunction with mechanisms like a bandwidth target multiplier (\S~\ref{sec:pacing_target}), enables \sqp to drain self-induced queues.
}{


The minimum packet transmission delay, $\Delta_{min}$, serves as a baseline for detecting self-inflicted network queuing. 
The window size for tracking $\Delta_{min}$ represents the duration of \sqp's memory of $\Delta_{min}$, which affects \sqp's self-induced queuing and throughput when competing with queue-building cross traffic. If a small, fixed window were used (e.g., 0.1-0.5~s), when self-induced queuing occurs, $\Delta_{\mathit{min}}$ could expire before \sqp can recover. While a larger, fixed window (e.g., 10-30~s) would aid recovery from self-induced queues by anchoring \sqp to the lowest one-way delay observed over the window, \sqp's bandwidth samples would be more sensitive to delay variations caused by the queue-building cross traffic, lowering \sqp's throughput share.

To balance these trade-offs, \sqp uses an adaptive window size of $2 \times \mathrm{sRTT}$~\cite{paxson2000computing}. This has two advantages. First, for self-induced queuing, \sqp's adaptive window grows quickly, and in conjunction with \sqp's bandwidth target multiplier mechanism (\S~\ref{sec:pacing_target}), enables \sqp to drain self-induced queues. Second, when competing queue-building flows build standing queues, $\Delta_{\mathit{min}}$ quickly increases in response, so that \sqp doesn't react to the competitor's standing queue. 
Since \sqp paces frames into the network faster than \sqp's current share, it can probe for more bandwidth when competing with queue-building flows, even if the combined link utilization of \sqp and the cross traffic is near 100\%. Together, this  enables \sqp to obtain a high throughput share when those queue-building flows (1) do not cause very high delays, and (2) have low queuing delay variation over periods of $2 \times \mathrm{sRTT}$ (\S~\ref{sec:eval:cross_traffic}).

}

While the role of \sqp's $\Delta_{min}$ mechanism is similar to \bbr's minRTT mechanism, \sqp does not need an explicit probing mechanism for $\Delta_{\mathit{min}}$ since it (1) increases the window size when self-induced queuing occurs, and (2) reduces the video bitrate to drain the self-induced queue, which provides organic stability.
While \bbr's explicit probing of the baseline network RTT is more accurate, the need to significantly limit packet transmissions for 200~ms makes this approach unsuitable for real-time interactive streaming media.
We evaluate the impact of the window size scaling parameter in \S~\ref{sec:min_owd_window_eval}.

\subsection{Bandwidth Estimate Update Rule}
\label{sec:bw_interp}
\sqp's bandwidth estimator processes noisy bandwidth samples measured by \sqp's bandwidth sampler to provide a smooth bandwidth, which is used to set the video encoder bitrate.
\sqp's update rule is inspired from past work on network utility optimization~\cite{kelly1998rate}, and is derived by optimizing a logarithmic reward for higher bandwidth estimates and a quadratic penalty for overestimating the bandwidth:
\begin{align}
    \label{eq:bw-est-expr}
    \max \log(1 + \alpha \cdot B) - \beta \cdot (B - e)^2
\end{align}
where $B$ is \sqp's bandwidth estimate, $\alpha$ is the reward weight for a higher bandwidth estimate, $\beta$ is the penalty for overestimating the bandwidth, and $e$ is a parameter derived from the bandwidth sample $S$, such that the function is maximized when $B=S$. Taking the derivative of this expression and evaluating the expression with $B$ set to the current estimate provides a gradient step towards the maxima.
Simplifying the derivative of the expression~\ref{eq:bw-est-expr}, and the constant expressions involving $\alpha$ and $\beta$, the update rule can be rewritten as
\begin{align}
    B^{\prime} = B + \delta \left(r \left(\frac{S}{B} - 1\right) - \left(\frac{B}{S}-1\right)\right)
    \label{eq:update_rule}
\end{align}
$B^{\prime}$ and $B$ are the updated and current estimates, $r$ is the reward weight for bandwidth utilization and $\delta$ is the step size and represents a trade-off between the smoothness of the bandwidth estimate and the convergence time under dynamic network conditions. \sqp empirically sets $\delta = 320~\si{kbps}$, and $r=0.25$.

\revised{
Intuitively, when a new bandwidth sample is much lower or much higher than the current bandwidth estimate, the update rule takes a larger step in the direction of the new bandwidth sample.
}
{
\sqp's target and pacing multiplier mechanisms (\S~\ref{sec:bw_samp_analysis}) work in conjunction with the update rule to improve \sqp's convergence to fairness (\S~\ref{fig:update_fairness_convergence}).
}

\subsection{Pacing and Target Multipliers}
\label{sec:pacing_target}
\sqp's design includes two key mechanisms for ensuring friendliness with other flows - instead of transmitting each frame as an uncontrolled burst at line rate, \sqp paces each frame at a multiple of the bandwidth estimate, and targets a slightly lower video bitrate than the samples (determined by a target multiplier).
Suppose \sqp is sharing a bottleneck link with a hypothetical elastic CCA~\cite{goyal2018elasticity-nimbus} that perfectly saturates the bottleneck link without inducing any queuing delay.
If \sqp transmitted frames as uncontrolled bursts, the elastic flow might not be able to insert any packets between \sqp's packets.
Thus, the bandwidth samples would match the link rate, and \sqp would starve the elastic flow by utilizing the entire link bandwidth.

\sqp paces each frame at a rate $P$, which is a multiple of the current bandwidth estimate, ie. $P = m \cdot B~(m > 1.0)$.
Thus, each frame is transmitted over $\frac{I}{m}$, where $I$ is the frame interval.
While pacing enables competing traffic to disperse \sqp's packets, \sqp's bandwidth samples would still be higher than the average rate it is sending at, and \sqp would eventually starve the other flow.
To avoid this problem, \sqp combines frame pacing with a bandwidth target multiplier mechanism.
\sqp multiplies bandwidth samples with a target multiplier $T < 1$ before calculating the bandwidth estimate.
\sqp's target multiplier serves three key roles: (1) it allows \sqp to drain any self-inflicted standing queues over time, (2) in conjunction with the pacing multiplier, it prevents \sqp from starving competing flows, and (3) enables multiple \sqp flows to converge to fairness.
We empirically set $m=2$ and $T=0.9$, and analyze the impact of other values of $T$ in Section \S~\ref{sec:bw_samp_analysis}.


\section{Analysis of \sqp Dynamics}
\subsection{Competing Flows}
\label{sec:bw_samp_analysis}

\begin{figure*}
\centering
\includegraphics[width=0.9\textwidth]{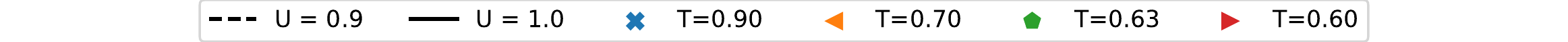}\\
\subcaptionbox{Theoretical utilization of available capacity.\label{fig:sqp_util_curve}}{\includegraphics[width=0.245\textwidth]{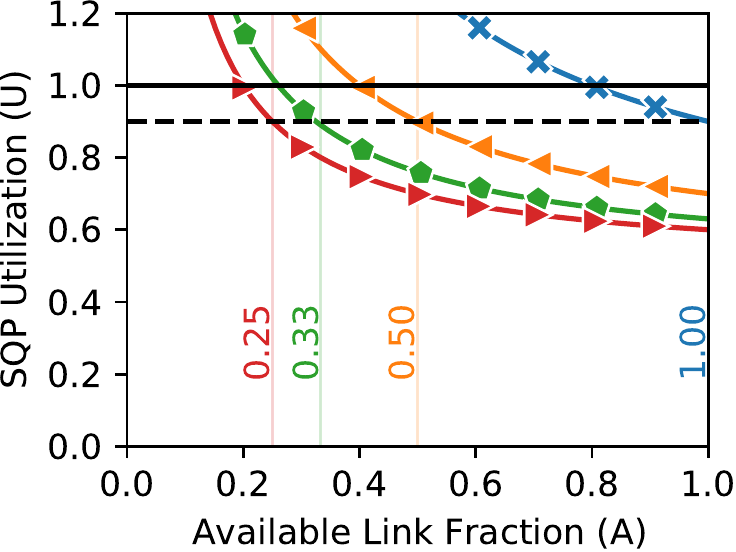}}%
\hfill
\subcaptionbox{Single \sqp flow throughput when competing with cross traffic. \label{fig:microbench:target_ct_share}}{\includegraphics[width=0.236\textwidth]{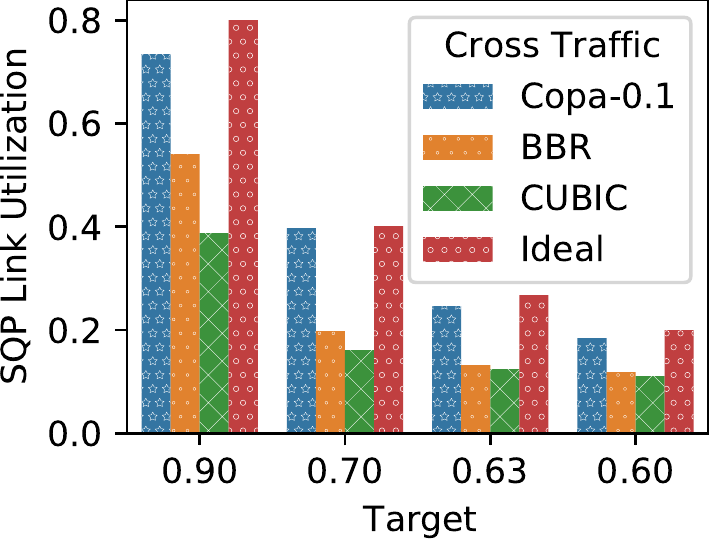}}%
\hfill
\subcaptionbox{P90 packet delay for \sqp flows sharing a bottleneck. \label{fig:microbench:target_delay}}{\includegraphics[width=0.234\textwidth]{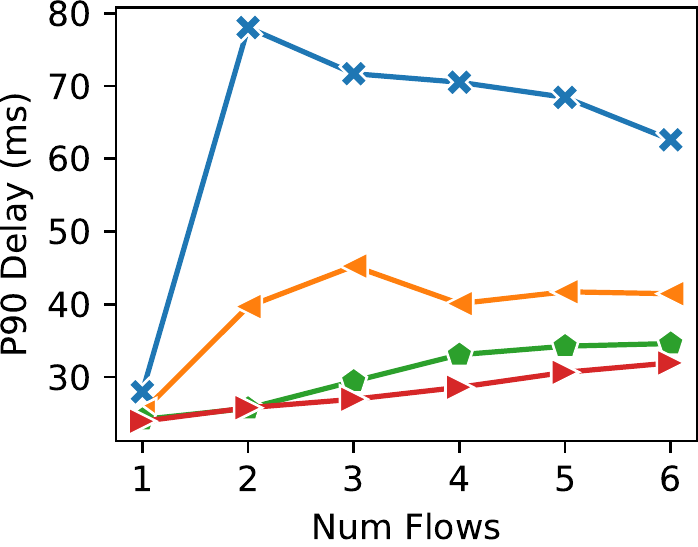}}%
\hfill
\subcaptionbox{Total link utilization for \sqp flows sharing a bottleneck. \label{fig:microbench:target_tput}}{\includegraphics[width=0.238\textwidth]{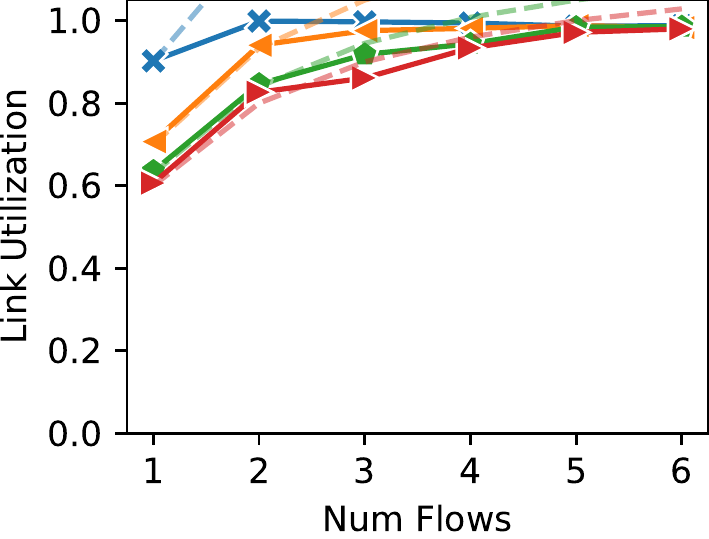}}%
\vspace{-0.2in}
\caption{Impact of the target multiplier on delay, link utilization and link share obtained under cross traffic for pacing multiplier $m=2$. Experimental results validate the theoretical analysis. In each case, the bottleneck link rate was 20 Mbps, the one-way delay in each direction was 20 ms and the bottleneck buffer size was 120 ms.}
\label{fig:microbench}
\end{figure*}

\revised{}{
\sqp's pacing multiplier ($m$) and bandwidth target multiplier ($T$) mechanisms provide important guarantees that prevent \sqp from starving other flows, and enable \sqp to achieve fairness when competing with other \sqp flows.
In this section, we derive \sqp's theoretical maximum share when competing with elastic flows, and the conditions under which \sqp achieves queue-free operation when competing with inelastic flows.
This analysis provides valuable insight into how \sqp's parameters can be tuned for application-specific performance requirements.
}

\sqp adds $B\cdot I$ bytes (i.e., the frame size, equal to the bandwidth estimate times the frame interval) to the bottleneck queue over a period $\frac{I}{m}$ (\S~\ref{sec:pacing_target}), during which a competing flow transmitting at a rate $R$ adds $\frac{R\cdot I}{m}$ bytes to the queue.
Thus, the time to drain the queue ($T_d$) is
\begin{align}
    T_d = \frac{B \cdot I + \frac{R \cdot I}{m}}{C}
    \label{eq:delivery_time_cross_traffic_disperse}
\end{align}
where $C$ is the link capacity. 
If the link is not being overutilized ($\Delta_{min}$ remains constant),
\begin{align}
T_d = R_{end}-R_{start} = R_{end}-S_{start}-\Delta_{min}
\label{eq:sqp_bwsample_time_delivery_time}
\end{align}
From Eq.~\ref{eq:sqp_bwsample_time_delivery_time} and Eq.~\ref{eq:bw_samp_owd}, \sqp's bandwidth sample ($S$) can be written as $S = \frac{B\cdot I}{T_d}$.
After substituting the value of $T_d$ from Eq.~\ref{eq:delivery_time_cross_traffic_disperse} and simplifying the equation, we get:
\begin{align}
S = \frac{C}{1 + \frac{R}{m \cdot B}}
\label{eq:sample_measured_from_pacing}
\end{align}

Note that we assume $m \cdot B + R > C$ (link is not severely underutilized), otherwise no queue will build up during the \sqp's pacing burst, and the bandwidth sample would simply be $m\cdot B$.
\sqp multiplies the bandwidth sample ($S$) with a target multiplier ($T$) before it is used to update the current bandwidth estimate using Eq.~\ref{eq:update_rule}.
Steady state occurs when the bandwidth estimate ($B$) is equal to the bandwidth target, ie. $B_T = S \cdot T$. 
Substituting $S$ from Eq. \eqref{eq:sample_measured_from_pacing}, we get:
\begin{align}
    B = C \cdot T - \frac{R}{m}
    \label{eq:steady_state_condition}
\end{align}

If $A = \frac{C-R}{C}$ is the fraction of the link capacity available for \sqp, and $U = \frac{B}{C-R}$ is \sqp's utilization of the of the available link capacity, Eq.~\ref{eq:steady_state_condition} can be re-written as
\begin{align}
    U =  \frac{m \cdot T + A - 1}{m \cdot A}
    \label{eq:util_avail}
\end{align}

This equation predicts \sqp's behavior in a variety of scenarios.
Figure~\ref{fig:sqp_util_curve} plots $U$ on the Y-axis as a function of $A$ on the X-axis for various target multipliers and for a pacing multiplier of 2.

\revised{}{
\textbf{Recovery From Self-Induced Queuing}
When \sqp is the only flow on a bottleneck link, the available link share $A = 1$ (right edge of Figure~\ref{fig:sqp_util_curve}).
This implies that \sqp will always underutilize the link slightly (specifically, it will use fraction $T$ of the total link capacity), which will result in standing queues getting drained over time.
The value of $T$ caps \sqp's maximum link utilization in the steady state, and determines how quickly \sqp will recover from self-induced standing queues.
In our evaluation and for \sqp's deployment in \company's AR streaming service (\S~\ref{sec:real_world_copa_ab}), we use a target multiplier of 0.9, which achieves good link utilization and is able to drain standing queues reasonably well.
}


\textbf{Inelastic Cross Traffic.}
When \sqp competes with an inelastic flow (transmitting at a fixed rate), the available link fraction ($A$) is fixed.
For \sqp to operate without any queuing, $U$ (\sqp's utilization of the available capacity)  must be less than 1.
Thus, the available bandwidth must be greater than the value at which the utilization curve crosses $U=1$ in Figure~\ref{fig:sqp_util_curve}.

For example, with a pacing multiplier of 2 and a target multiplier of 0.9, \sqp requires at least 80\% of the link to be available so that it can consistently maintain a slight underutilization of the link.
When less than $80\%$ of the link is available, \sqp will tend to overutilize its share and cause queuing. 
While \sqp's initial window size for tracking $\Delta_{min}$ (~\ref{sec:min_owd_window}) may not be large enough for \sqp to completely drain the self-induced queue, the increase in the RTT due to queuing will eventually cause the window to grow to a size that is large enough to stabilize \sqp's queuing.
While a smaller target value would enable \sqp to operate without queuing for lower values of $A$, it would sacrifice link utilization when there are no competing flows.

\textbf{Elastic Cross Traffic.} 
The minimum value of $A$ for queue-free operation of \sqp when competing with inelastic flows is also the  maximum bound for \sqp's share of the throughput when it is competing with elastic traffic.
When \sqp is not using its entire share ($U < 1$), the elastic flow will increase its own share since the link is underutilized.
This reduces the available link share for \sqp, moving the operating point to the left in Figure~\ref{fig:sqp_util_curve} until the entire link is utilized ($U = 1$).
If \sqp is over-utilizing its share, the elastic flow will decrease its own share and move the operating point to the right until the link is no longer being over-utilized.

This is an upper-bound of \sqp's share. Non-ideal elastic flows can cause queuing delays that will cause \sqp to increase its one-way delay tracking window, whch in turn will make the bandwidth samples more sensitive to delay variations caused by the cross traffic.
Figure~\ref{fig:microbench:target_ct_share} shows the share of a single \sqp flow competing with various elastic flows, for different target multipliers.
\copa-0.1 closely resembles an ideal elastic flow which does not cause queuing and has low delay variation. 
Hence, \sqp's share (shown in blue) is close to the theoretical maximum (shown in red).
With \bbr and \cubic, \sqp's share is less than the theoretical maximum since the higher delays induced by \bbr and \cubic make \sqp more reactive to queuing delay variations.

\subsection{Intra-protocol Dynamics and Fairness}
\label{sec:intraprotocol_fairness}
From the analysis in \S~\ref{sec:bw_samp_analysis}, we can also infer the number of \sqp flows that can operate without queuing on a shared bottleneck, with some caveats.
The underlying assumption in \S~\ref{sec:bw_samp_analysis} is that packet arrivals at the bottleneck are evenly spaced.
The analysis also holds in the case of Poisson arrivals since the bandwidth samples are smoothed out by the update rule (\S~\ref{sec:bw_interp}).
Multiple \sqp flows transmit frames as regularly spaced bursts, and thus, the packet dispersion observed by one \sqp flow depends on how its frames align with the frames of the other flows.
If the two flows that are sharing the bottleneck have perfectly aligned frame intervals, each flow will observe exactly half of the link rate, and they will operate without queuing since $T<1$.
If the frame intervals are offset exactly by $I/2$ (when pacing at 2X), each flow will see the full link bandwidth until link overutilization triggers \sqp's transient delay recovery mechanism.
When the intervals are offset by $I/4$, the packet dispersion is the same as the dispersion caused by a uniform flow.
Note that this is only a concern if there are very few \sqp flows, and the applications have perfectly timed frames.
As the number of \sqp flows increase, the aggregate traffic pattern gets smoothed out.

Figures~\ref{fig:microbench:target_delay} and~\ref{fig:microbench:target_tput} show the 90th percentile delay and the total link utilization respectively on the Y-axis as a function of the number of flows for various target multipliers.
Figure~\ref{fig:microbench:target_tput} plots the theoretical link utilization of multiple \sqp flows using dashed lines.
The pacing multiplier was set to 2.0 for all runs.
To avoid the impact of frame alignment in our experiment, we incorporate 1 ms of jitter into the frame generation timing\footnote{Incorporating 1ms of sub-frame jitter into an application's frame rendering will have minimal impact on video smoothness}.
When $T=0.9$, a single \sqp flow in isolation maintains low delay and utilizes 90\% of the link; two or more \sqp flows fully utilize the link and stabilize at a slightly higher delay (similar to \sqp's behavior with inelastic cross-traffic, \S~\ref{sec:bw_samp_analysis}).
Reducing the target value reduces the steady state queuing delay, with the trade-off that an isolated \sqp flow will have lower link utilization  (Figure~\ref{fig:microbench:target_tput}) and will obtain less throughput share when competing with elastic flows (\S~\ref{sec:bw_samp_analysis}).

While \sqp can be adapted to use more sophisticated mechanisms like dynamic frame timing alignment across \sqp flows and dynamically lowering the target and pacing multipliers when the presence of multiple \sqp flows is detected, we defer this to future work and only evaluate the base \sqp algorithm with a fixed target multiplier $T=0.9$ and a fixed pacing multiplier $m=2$ in \S~\ref{sec:evaluation}.

\begin{figure}
\centering
\subcaptionbox{Raw sample update\label{fig:raw_sample_update}}{\includegraphics[width=0.42\columnwidth]{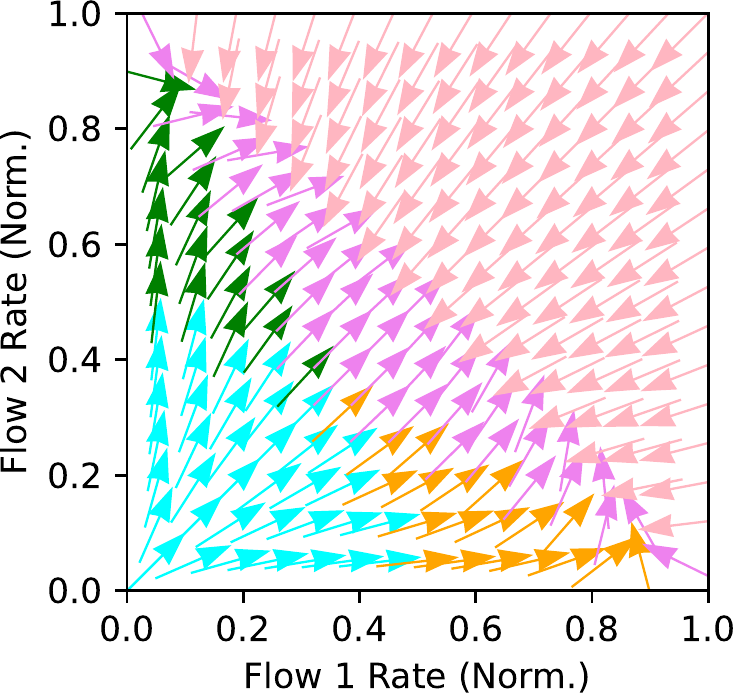}}%
\hfill
\subcaptionbox{Using update rule\label{fig:update_rule_update}}{\includegraphics[width=0.42\columnwidth]{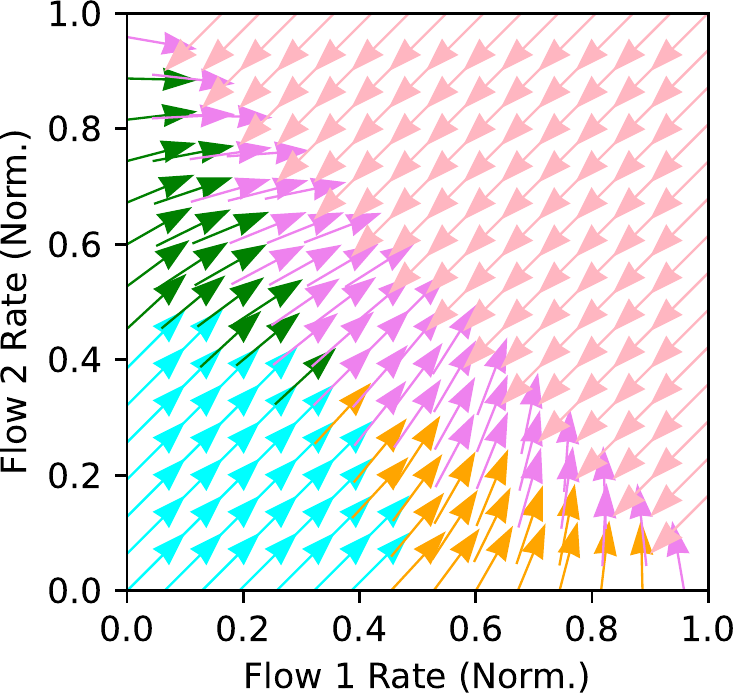}}%
\vspace{-0.1in}
\caption{Vector field showing bandwidth update steps for different starting states for two competing flows. \sqp's update rule significantly speeds up convergence to fairness.}
\label{fig:update_fairness_convergence}
\end{figure}
\revised{
\textbf{Fairness.} 
The convex nature of the utilization curves shown in Figure~\ref{fig:sqp_util_curve} in conjunction with \sqp's bandwidth update rule (\S~\ref{sec:bw_interp}) ensures fast convergence to fairness among multiple \sqp flows, and avoids the impact of the late-comer effect on fairness between multiple flows (Figures~\ref{fig:eval:10flow:60mbit_throughput_delay} and~\ref{fig:eval:fairness:fairness_30frames}).
}
{
\textbf{Fairness.}
When competing with other \sqp flows, there are two key mechanisms that enable \sqp to converge to fairness: \sqp's pacing-based bandwidth probing (\S~\ref{sec:bw_sampling}), and \sqp's logarithmic utility-based bandwidth smoothing (\S~\ref{sec:bw_interp}).

Let's consider a scenario where \sqp is not using bandwidth smoothing, and directly updates its bandwidth estimate according to the sample.
When overutilization occurs, each flow observes a common delay signal, and hence the bandwidth is reduced by a multiplicative factor.
For various values of each flow's initial rate, we compute the update step as $S\times T - B$, where $S$ is computed using Eq.~\ref{eq:sample_measured_from_pacing} (average case behavior with randomized frame alignment, \S~\ref{sec:bw_samp_analysis}).
When the link is severely underutilized by a flow (the pacing burst of \sqp does not cause queuing - see \S~\ref{sec:bw_samp_analysis}), the update step is $2\times B - B = B$.
These update steps are shown in Figure~\ref{fig:raw_sample_update} as arrows, where the tail of the arrow is anchored at the initial condition, and the length of the arrow is proportional to the step size.
In the cyan region, neither of the flows cause queuing due to their pacing burst, and hence, the rates undergo a multiplicative increase (direction of increase passes through the origin on the graph).
In the purple region, both flows cause temporary queuing due to their pacing burst, and the slower flow increases its rate more than the faster flow (whose increase is sublinear).
The green and orange regions depict a region of transition, where only one of the flows observes pacing-induced queuing.
Thus, while \sqp will undergo multiplicative increase when the link is severely underutilized, as the link utilization increases, the increases become sublinear. A downside is that convergence is slow when the link is severely underutilized - a faster flow will increase it's rate more quickly as compared to a slower flow initially due to the multiplicative increase. We solve this using \sqp's bandwidth update rule (\S~\ref{sec:bw_interp}), which significantly speeds up convergence to fairness.

In Figure~\ref{fig:update_rule_update}, we compute the update steps by incorporating \sqp's logarithmic utility-based bandwidth update rule. In this case, \sqp undergoes linear increase when the link is severely underutilized, sublinear increase when the link is close to being fully utilized, and linear decrease when the link is overutilized.
Thus, \sqp converges to fairness (similar to AIMD).
The linear increase speeds up convergence for multiple \sqp flows from a severely under-utilized state. The linear decrease makes \sqp's throughput stable when competing with queue-building flows, since it does not react as fast as multiplicative decrease.
\sqp's bandwidth update rule also ensures that the updates are proportional to the difference relative to the current estimate, as opposed to fixed-size steps (e.g., additive increase in \cubic) or velocity-based mechanisms (e.g., Copa). We evaluate \sqp's fairness in \S~\ref{sec:eval:fairness}.
\srini{need to provide forward pointer to fairness measurement}\dd{Done}

}

\subsection{Adaptive Min One-way Delay Tracking}
\label{sec:min_owd_window_eval}
\begin{figure}
\centering
\subcaptionbox{P90 packet delay for multiple \sqp flows.\label{fig:rtt_mx_sqp_delay}}{\includegraphics[width=0.49\columnwidth]{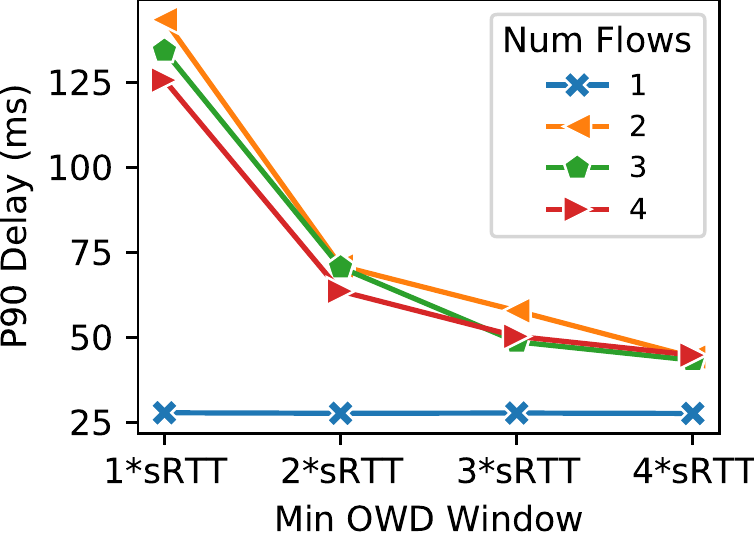}}%
\hfill
\subcaptionbox{Throughput of 1 \sqp flow in the presence of cross traffic. \label{fig:rtt_mx_ct_tput}}{\includegraphics[width=0.465\columnwidth]{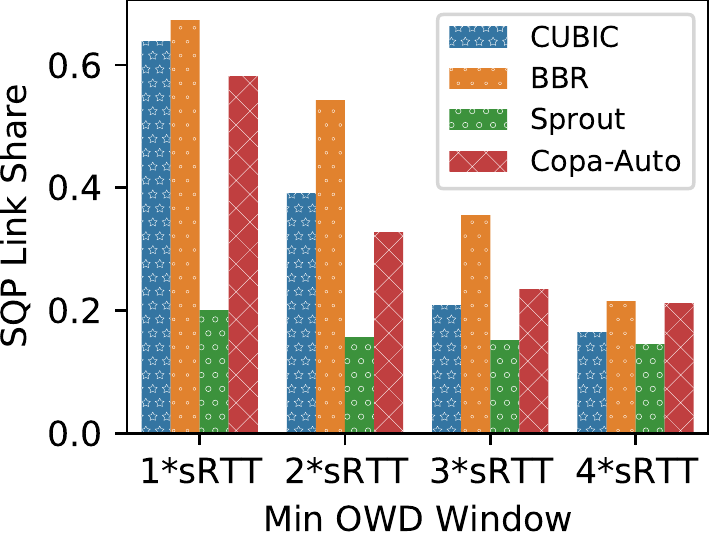}}%
\caption{Impact of the min one-way delay multiplier on frame delay and throughput when competing with other flows. The bottleneck setup is the same as Figure~\ref{fig:microbench}, and $T=0.9, m=2$.}
\label{fig:minowd_benchmark}
\end{figure}

\sqp's adaptive min one-way delay window is a key mechanism that enables \sqp to recover from network overutilization.
Recall that \sqp's window scales with the currently observed sRTT (\S~\ref{sec:min_owd_window}).
A larger window speeds up recovery from queuing caused by overutilization, but results in poor performance when competing with queue-building cross traffic.
Different multipliers are evaluated in Figure~\ref{fig:minowd_benchmark}.
With $T=0.9$ and $m=2$, more than one \sqp flows sharing a bottleneck require a larger $\Delta_{min}$ window to stabilize. 
A multiplier of 2 results in acceptable level of steady state queuing (nearly as low as $3\times$ and $4\times$), while achieving reasonable throughput in the presence of queue-building cross traffic like \cubic and \bbr. 
\sqp competing with \sprout is also shown as a worst case example; \sprout causes significant delay variation due to its bursty traffic pattern, causing \sqp to achieve low throughput.

\revised{
\subsection{Discussion}
\textbf{Shallow Buffers.} 
For the target workload of interactive video with $I=16.66~\si{\milli\second}$ (60FPS) and a pacing multiplier $m=2$,
\sqp's pacing-based bandwidth probing only requires approximately 8 ms of packet buffer at the bottleneck link to be able to handle the burst for each frame.
Typical last-mile network links like DOCSIS, cellular, and Wi-Fi links have much larger packet buffers~\cite{gettys2011bufferbloat}.

When \sqp competes with other flows (vs. \sqp, inelastic flows), \sqp may require a higher level of queuing to achieve its approximate fair share of bandwidth.
For handling such scenarios in the presence of shallow buffers, \sqp can simply respond to persistent packet loss by reducing the pacing multiplier or the target rate, since both these steps significantly reduce the amount of queuing required by \sqp.
Alternatively, the TFRC~\cite{padhye1999model} rate control equation can be integrated into \sqp to achieve good performance and approximate fairness in these scenarios.

\textbf{Minimum Bitrate.} \sqp relies on transmitting multiple packets per frame to estimate the link bandwidth, and thus requires at least 2 packets per frame (approximately 1 Mbps for 1200 byte packets at 60 FPS).
In such scenarios, the frame rate can be reduced or smaller packets can be sent so that \sqp can effectively probe for more bandwidth.
}
{}
\section{Evaluation}
\label{sec:evaluation}

\begin{figure*}
    \centering
    \includegraphics[width=0.7\textwidth]{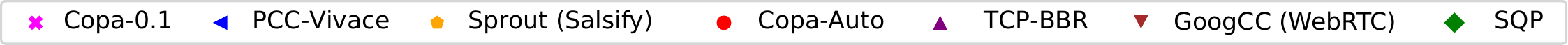}\\
    \subcaptionbox{Send rate timeseries.\label{fig:appendix:sqp_timeseries}}{
    \includegraphics[width=0.3\textwidth]{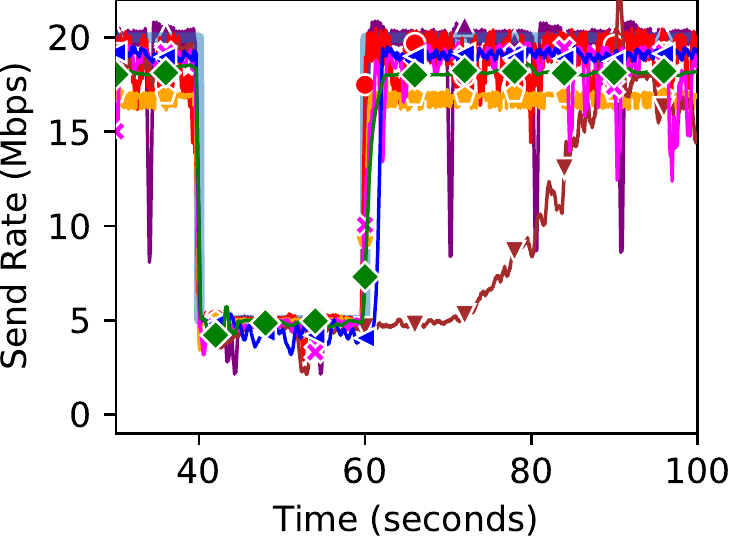}
    }%
    \hfill
    \subcaptionbox{Packet delay timeseries.\label{fig:appendix:sqp_delay}}{
    \includegraphics[width=0.3\textwidth]{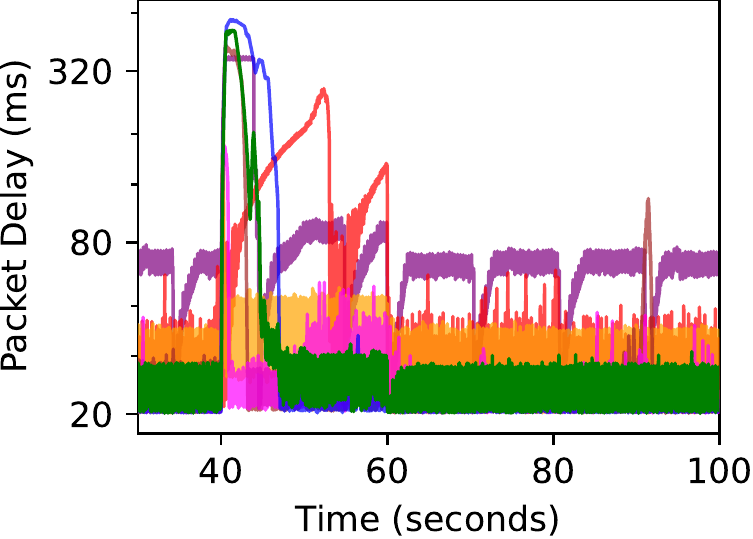}
    }%
    \hfill
    \subcaptionbox{Packet delay timeseries (zoomed in).\label{fig:appendix:sqp_delay_zoomed}}{
    \includegraphics[width=0.3\textwidth]{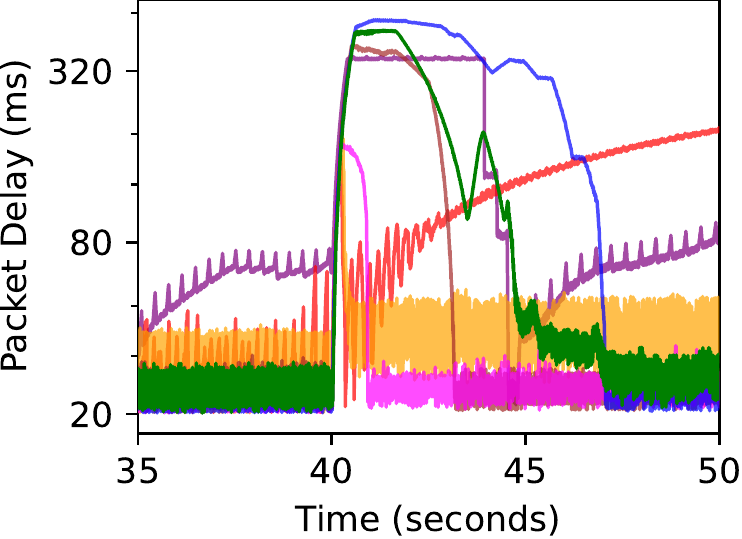}
    }%
    \caption{Congestion control performance on a variable link (link bandwidth shown as a shaded light blue line).}
    \label{fig:appendix:experiment_timeseries}
\end{figure*}

\sqp's evaluation has three broad themes. 
\S~\ref{sec:eval:real-world-trace} evaluates \sqp's performance on a large set of calibrated emulated links modeled after real-world network traces obtained from \company's game streaming service.
\S\S~\ref{sec:eval:cross_traffic}-\ref{sec:encoder_variation} evaluate \sqp's throughput when competing with cross traffic, impact of shallow buffers, fairness, and bandwidth probing in application-limited scenarios.
In \S~\ref{sec:real_world_copa_ab}, we compare \sqp and \copa (without mode switching) in the real world on \company's AR streaming service.
In this section we compare \sqp's performance to recently proposed high performance low latency algorithms like \pcc~\cite{dong2015pcc}, \copa~\cite{arun2018copa} (with and without mode switching), \vivace~\cite{dong2018pcc} and \sprout~\cite{winstein2013stochastic}, traditional queue-building algorithms like TCP-\cubic~\cite{ha2008cubic} and TCP-\bbr~\cite{cardwell2016bbr}, and WebRTC (using GoogCC as CCA), an end-to-end low-latency streaming solution.

\subsection{Emulation Setup}
\label{sec:eval:setup}
We use the Pantheon~\cite{yan2018pantheon} congestion control testbed, which works well for links under $\SI{100}{Mbps}$.
For the baselines, we use the implementations available on Pantheon.
These include kernel-space (Linux) implementations of \cubic and \bbr-v1~\cite{bbr-v1-impl} (iperf3~\cite{tirumala1999iperf}), user-space implementations of \pcc~\cite{pcc-impl}, \vivace~\cite{vivace-impl}, \copa~\cite{copa-impl}, and \sprout, and Chromium's version of WebRTC (with GoogCC, max bitrate changed to 50 Mbps from 2 Mbps).
Additionally, we evaluate the \copa algorithm with a fixed delta ($\delta=0.1$).
We implement additional functionality in Pantheon, including flow-specific RTTs, start and stop times, and testing of heterogeneous CCAs sharing a link.
For experiments with fixed bandwidth links, we choose a queue size of 10 packets / Mbps ($\approx 120 \si{\milli\second}$ for 20 Mbps) and the drop-tail queuing discipline.
We fix $T=0.9$ and $m=2.0$ for \sqp.

\subsection{Metrics}
\label{sec:eval:metrics}
While metrics like average throughput and packet delay are sufficient for evaluating a general purpose congestion control algorithm, they do not accurately reflect the impact on quality-of-experience (QoE) of a low-latency streaming application that is using a particular congestion control algorithm~\cite{liu2016streaming}.
To evaluate how a CCA affects the QoE of low-latency streaming, we need metrics that quantify properties like video bitrate and frame delay.

After an experiment is run, Pantheon generates detailed packet traces with the timestamps of packets entering and leaving the bottleneck.
We compute a windowed rate from the ingress packet traces, which serves as a baseline for the video bitrate.
For a time slot $t$, the frame size $F(t)$ is:
\begin{align}
    F(t) = max\left(\frac{S(t, t+n \cdot I) - p}{n}, \; S(t, t+I) - p,\; 0\right)
\end{align}
\noindent
where $p$ denotes the pending unsent bytes from previous frames, $I$ is the frame interval, $S(t_1, t_2)$ is the number of bytes sent by an algorithm between $t_1$ and $t_2$ and $n$ is the window size in number of frames used for smoothing. This ensures that none of the bytes the algorithm sent in a particular interval are wasted (maximum utilization).

To quantify video frame delay, we simulate the transmission of the frames to measure the end-to-end frame delay. For zero size frames, we assume that the delay of the frame is the time until the next frame. 
The choice of $n$ limits the worst case sender-side queuing delay to $n$ frames, which can occur when an algorithm sends a burst of packets during the $n^{th}$ frame slot after a  quiescence period of $n-1$ frames.

\subsection{Simple Variable Bandwidth Link}
\label{sec:appendix:timeseries}
\srini{would be good to have a backward reference to paper}\dd{It is there .. (same as the experiment described in .. )}
We evaluated \sqp on a link that runs at 20 Mbps for 40 seconds, drops to 5 Mbps for 20 seconds, and then recovers back to 20 Mbps (same as the experiment described in \S~\ref{sec:prelim_study}).
The throughput is shown in Figure~\ref{fig:appendix:sqp_timeseries}, and the delay is shown in Figure~\ref{fig:appendix:sqp_delay}, with a zoomed version of the delay in Figure~\ref{fig:appendix:sqp_delay_zoomed}.
\sqp quickly probes for bandwidth after the link rate increases ($T=60$), and is able to maintain consistent, low delay when the link conditions are stable.
When the link rate drops, \sqp's recovery is faster than \pcc-Vivace, and as fast as \bbr.
While GoogCC's recovery is slightly faster, it takes a very long time compared to \sqp in order to ramp up once the link rate increases back to 20.

\subsection{Real-world Wireless Traces}
\label{sec:eval:real-world-trace}
To evaluate \sqp's performance on links with variable bandwidth, delay jitter and packet aggregation, we obtained 100 LTE and 100 Wi-Fi throughput and delay traces from a cloud gaming service.
Each network trace was converted to a MahiMahi trace using packet aggregation to emulate the delay variation, and the link delay was set to the minimum RTT for each trace.

\begin{figure}
    \centering
    \includegraphics[width=1.0\columnwidth]{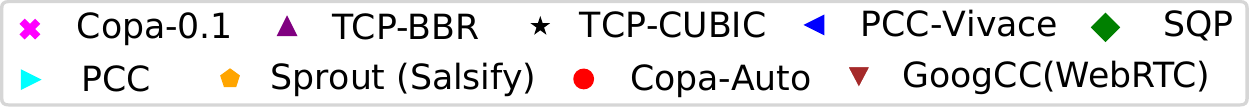}
    \subcaptionbox{Throughput timeseries for a sample Wi-Fi trace. \label{fig:eval:trace:wifi:bwtrace}}[0.48\columnwidth]{\includegraphics[width=0.49\columnwidth]{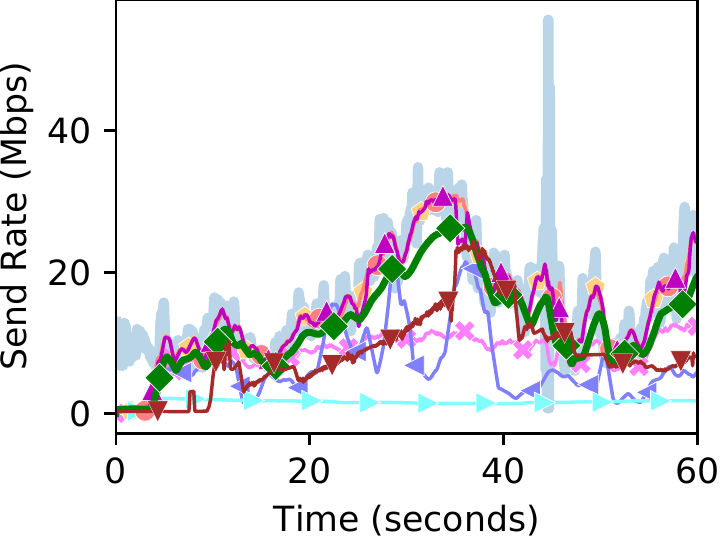}}%
    \hfill
    \subcaptionbox{Packet delay timeseries for the trace shown in \ref{fig:eval:trace:wifi:bwtrace}. \label{fig:eval:trace:wifi:delaytrace}}[0.48\columnwidth]{\includegraphics[width=0.503\columnwidth]{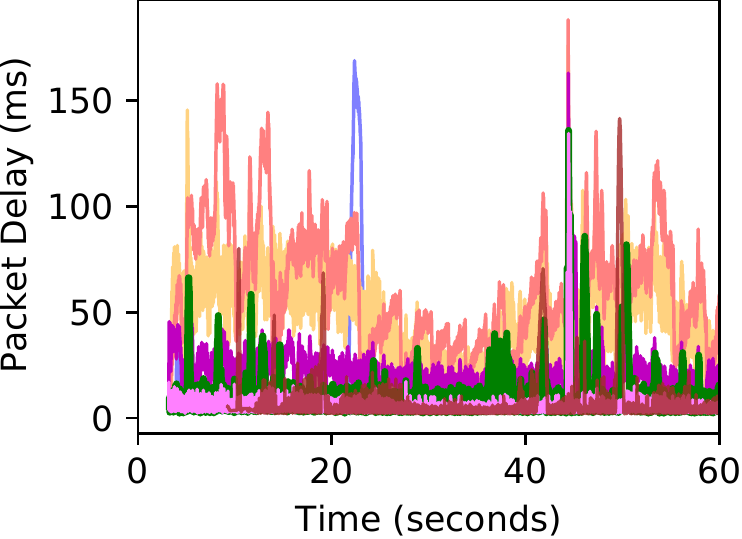}}
    \caption{Performance of various CCAs on a sample Wi-Fi network trace, with the bottleneck buffer size set to 200 packets. SQP rapidly adapts to the variations in the link bandwidth, and achieves low packet queueing delay.}
\end{figure}

Figures~\ref{fig:eval:trace:wifi:bwtrace} and~\ref{fig:eval:trace:wifi:delaytrace} show the throughput and delay of a single flow operating on a representative Wi-Fi trace.
The thick gray line represents the link bandwidth.
\sqp achieves high link utilization and can effectively track the changes in the link bandwidth while maintaining low delay.
While \copa-Auto, \sprout, and \bbr achieve high link utilization, they incur a high delay penalty.
\webrtc, \pcc and \vivace are unable to adapt to rapid changes in the link bandwidth, resulting in severe link underutilization and occasional delay spikes (e.g., \vivace at T=22s).

\begin{figure}
    \centering
    \includegraphics[width=\columnwidth]{figures/legend_sigcomm_sqp.pdf}
    \subcaptionbox{Performance across 100 real-world Wi-Fi traces. \label{fig:eval:trace:wifi:agg}}[0.48\columnwidth]{\includegraphics[width=0.48\columnwidth]{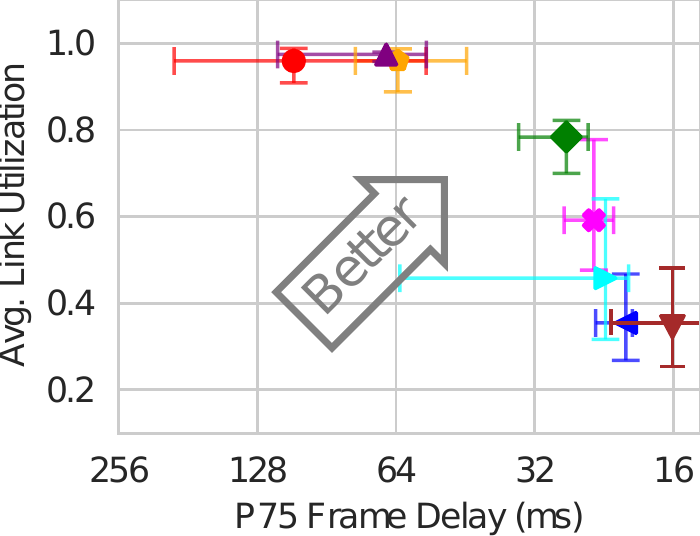}}%
    \hfill
    \subcaptionbox{Performance across 100 real-world LTE traces.\label{appendix_fig:eval:trace:lte:agg}}{\includegraphics[width=0.48\columnwidth]{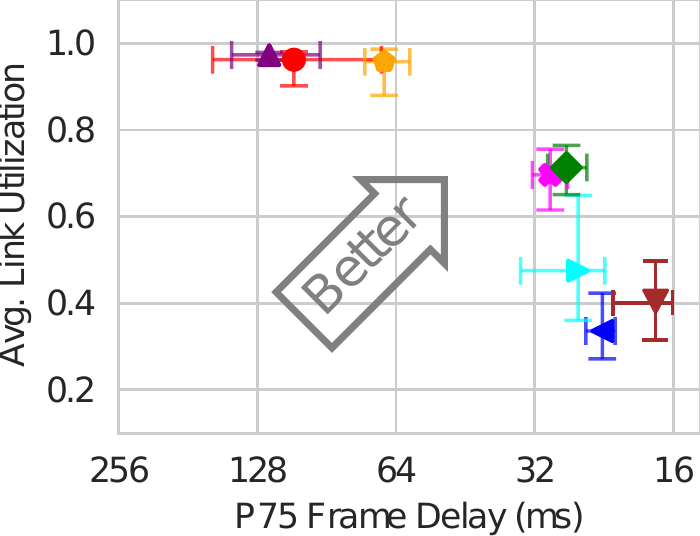}}
    \caption{\sqp's performance over emulated real-world wireless network traces. The bottleneck buffer size was set to 200 packets. In Figures~\ref{fig:eval:trace:wifi:agg} and~\ref{appendix_fig:eval:trace:lte:agg}, the markers depict the median across traces and the whiskers depict the $25^{\rm th}$ and $75^{\rm th}$ percentiles. \label{fig:eval:real_world_traces}}
\end{figure}

The aggregated results for the Wi-Fi traces are shown in Figure~\ref{fig:eval:trace:wifi:agg}.
Across all Wi-Fi traces, \sqp achieves 78\% average link utilization compared to 46\%, 35\% and 59\% for \pcc, \vivace and \copa-0.1 respectively while only incurring 4-8 ms higher delay.
While \cubic, \bbr, \sprout, and \copa-Auto achieve higher link utilization, this is at a cost of significantly higher delay (130-342\% higher).

\begin{figure*}
    \centering
    \includegraphics[width=0.9\textwidth]{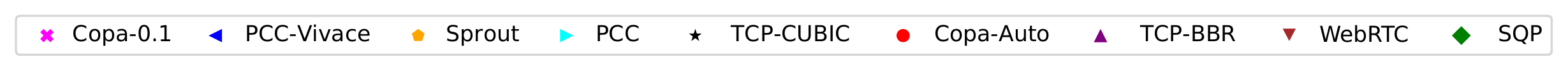}
    \subcaptionbox{Streaming performance when a \bbr flow starts after the primary flow has reached steady state.\label{fig:eval:cross_traffic_bbr}}{\includegraphics[width=0.49\columnwidth]{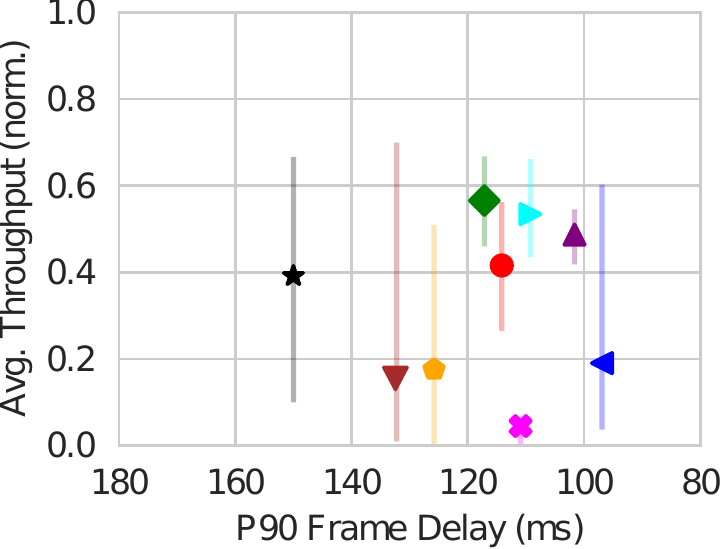}}%
    \hfill
    \subcaptionbox{Streaming performance when a \cubic flow starts after the primary flow has reached steady state.\label{fig:app:cross_traffic_cubic}}{\includegraphics[width=0.49\columnwidth]{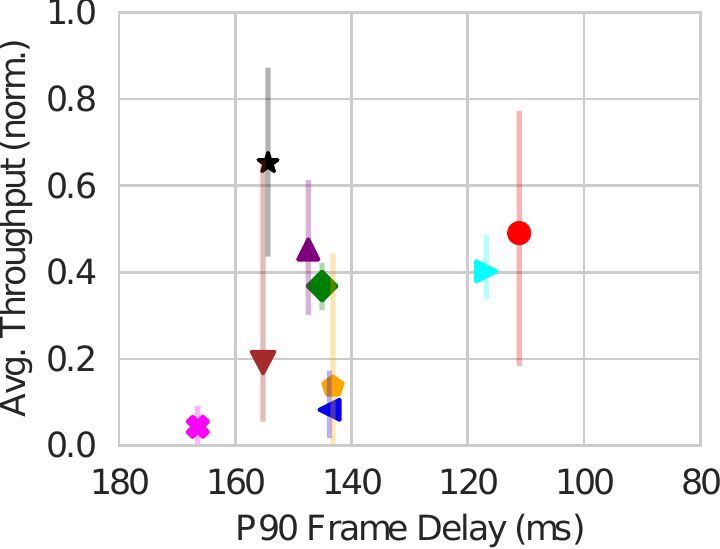}}%
    \hfill
    \subcaptionbox{Streaming performance when a CCA starts after a \bbr flow has reached steady state.\label{fig:eval:bbr_rev}}{\includegraphics[width=0.49\columnwidth]{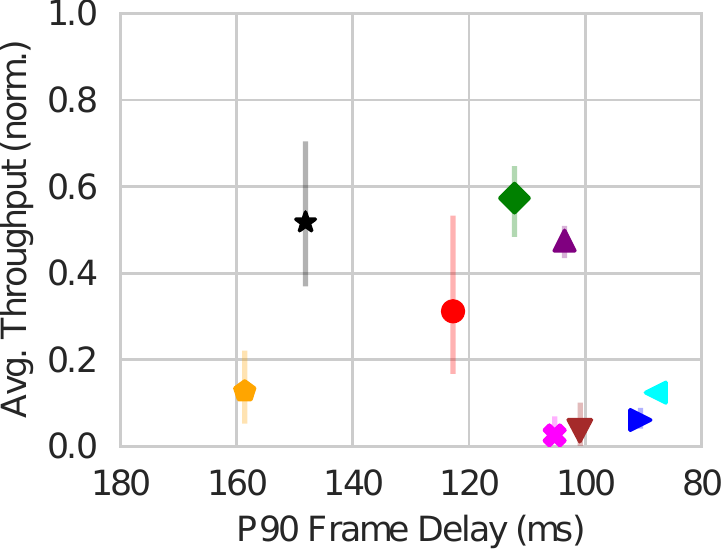}}%
    \hfill
    \subcaptionbox{Streaming performance when a CCA starts after a \cubic flow has reached steady state.\label{fig:eval:cubic_rev}}{\includegraphics[width=0.49\columnwidth]{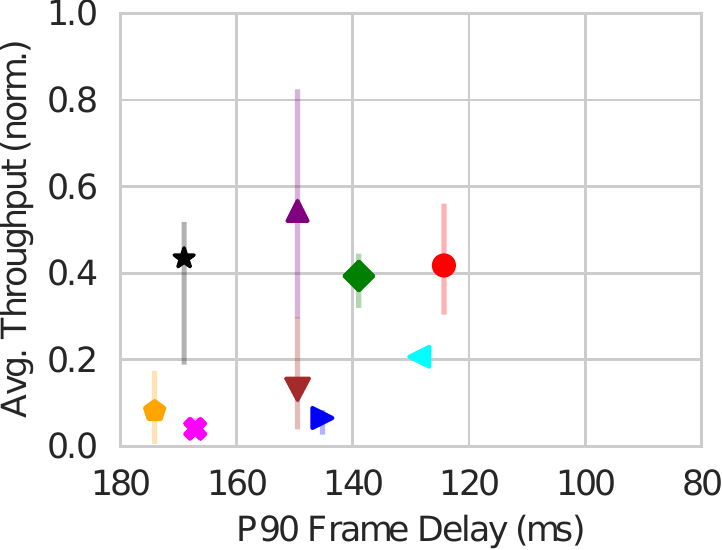}}%
    \hfill
    \caption{ CCA performance when competing with queue-building cross traffic. The error bars mark the P10 and P90 simulated frame bitrates (\S~\ref{sec:eval:metrics}).}
\end{figure*}

Figure~\ref{appendix_fig:eval:trace:lte:agg} shows the performance various CCAs across 100 real-world LTE traces. \sqp and Copa-0.1 have good throughput and delay characteristics, whereas other CCAs either have very high delays or insufficient throughput.

\subsection{Competing with Queue-building Flows}

\label{sec:eval:cross_traffic}


Next, we evaluate the ability of various congestion control algorithms to support stable video bitrates in the presence of queue-building cross traffic.
We ran the experiment for 60 seconds on a 20 Mbps bottleneck link with 120 ms of packet buffer, and a baseline RTT of 40 ms, where each algorithm is run for 10 seconds before the cross traffic is introduced.
Figure~\ref{fig:eval:cross_traffic_bbr} shows the average normalized throughput and P10-P90 spread of the windowed bitrate for each algorithm versus the P90 simulated frame delay after a \bbr flow is introduced.
Figure~\ref{fig:eval:bbr_rev} shows the average normalized throughput versus the P90 simulated frame delay when the CCA being tested starts 10 seconds after a BBR flow is already running on the link.
We ran similar experiments with \cubic as the cross traffic, and the results are shown in Figures~\ref{fig:app:cross_traffic_cubic} and~\ref{fig:eval:cubic_rev}.

\sqp is able to achieve high and stable throughput due to \sqp's bandwidth sampling mechanism (\S~\ref{sec:bw_sampling}) and the use of a dynamic min-oneway delay window size (\S~\ref{sec:min_owd_window}). 
While \pcc performs well when it starts before the competing traffic is introduced on the link, \pcc's normalized throughput is less than 0.2 when it starts on a link that already has a \bbr or \cubic flow running on it. \srini{would it be better to have fig 16d here instead? don't need to make this caveat then. }
GoogCC's slower start affects its throughput, with things improving slightly if the \cubic flow starts after 20s (App.~\ref{sec:appendix:webrtc_startup}), and it is also suffers from the latecomer effect.
\vivace, \copa-0.1 and \sprout are unable to maintain high throughput in all the cases.
While \copa-Auto has good average throughput, its performance is unstable at the frame timescale, which is evident by the spread between the P10 and P90 bitrate.

\revised{
}
{

\subsection{Shallow Buffers}
\label{sec:eval:shallow_buf}

\begin{figure}[htbp]
    \centering
    \includegraphics[width=\columnwidth]{figures/legend_sigcomm_sqp.pdf}\\
    \subcaptionbox{Ingress and Egress rate vs. Buffer Depth.\label{fig:eval:qdepth_rate}}{\includegraphics[width=0.48\columnwidth]{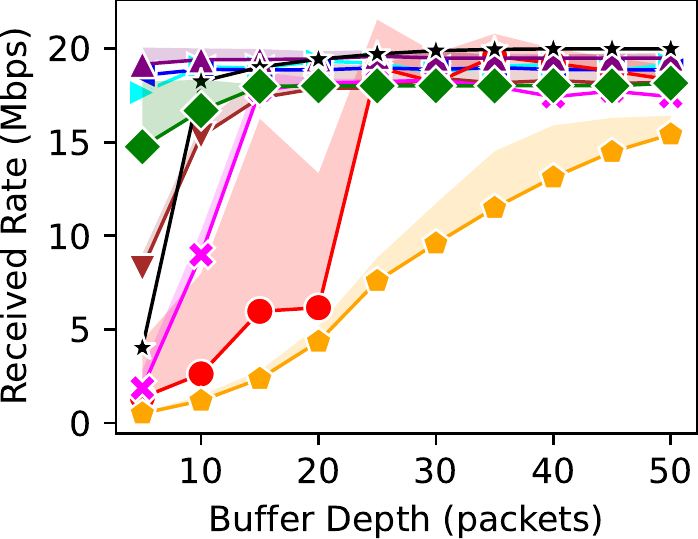}}%
    \hfill
    \subcaptionbox{Packet Loss rate vs. Buffer Depth.
    \label{fig:qdepth_loss_rate}}{\includegraphics[width=0.48\columnwidth]{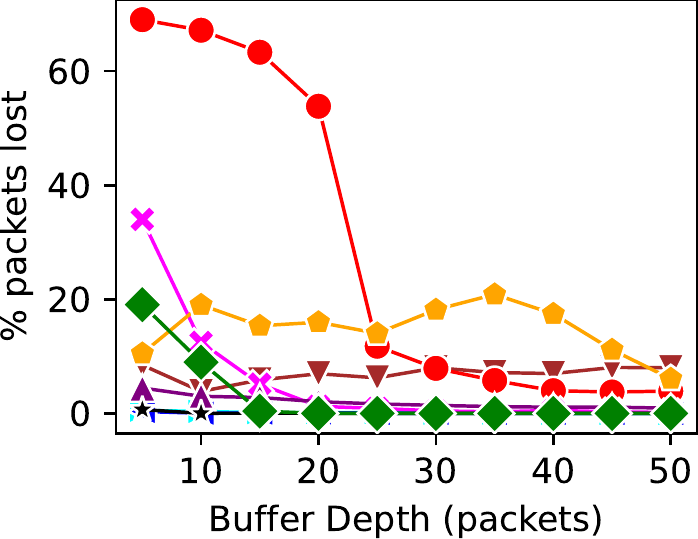}}
    \caption{Performance impact of shallow buffers on a 20 Mbps, 40ms RTT link.}
    \label{fig:eval:qdepth}
\end{figure}
For the target workload of interactive video with $I=16.66~\si{\milli\second}$ (60FPS) and a pacing multiplier $m=2$,
\sqp's pacing-based bandwidth probing only requires approximately 8 ms of packet buffer at the bottleneck link to be able to handle the burst for each frame.
The lines in Figure~\ref{fig:eval:qdepth_rate} show the link egress rate for various CCAs for different buffer sizes,
and the shaded regions denote the rate of loss (i.e., the top of the shaded region is the link ingress rate).
The loss rate is also shown in Figure~\ref{fig:qdepth_loss_rate}.
\sqp achieves its maximum throughput with a buffer of 15 packets or more, which corresponds to ~8 ms of queuing on a 20 Mbps link. 
If the buffer size is smaller than 15 packets, \sqp transmits at 18 Mbps ($T=0.9$ fraction of the link capacity), but the packets that correspond to the tail end of each frame are lost.
Copa-Auto, Sprout and GoogCC require larger buffers, whereas \bbr (\~{}4\% loss with a 5 packet buffer), and both \pcc versions (<1\% loss with a 5 packet buffer) excel at handling shallow buffers.
Typical last-mile network links like DOCSIS, cellular, and Wi-Fi links have much larger packet buffers~\cite{gettys2011bufferbloat}.
When \sqp competes with other flows (vs. \sqp, inelastic flows), \sqp may require a higher level of queuing to stabilize.
Dynamic pacing and target mechanisms are required to handle such scenarios, and we leave that for future work.

\textbf{Discussion: }
While \sqp causes sub-frame queuing since it paces each frame at 2X of the bandwidth estimate, this queuing is limited to a maximum of 8 ms (~14 packets for 20 Mbps). Hence, for buffer sizes of 15 packets and above, \sqp has exactly zero loss. \sprout on the other hand has 10-20 \% loss for buffer sizes all the way up to 50 packets, and more than 60\% of packets sent by Copa-Auto are lost for buffer sizes lower than 20 packets. GoogCC (WebRTC) has around 10\% packet loss for the entire range of buffer sizes evaluated here, which may be due to WebRTC sending FEC packets in response to the loss observed.

\subsection{Short Timescale Variations}

\begin{figure}[htbp]
    \centering
    \includegraphics[width=\columnwidth]{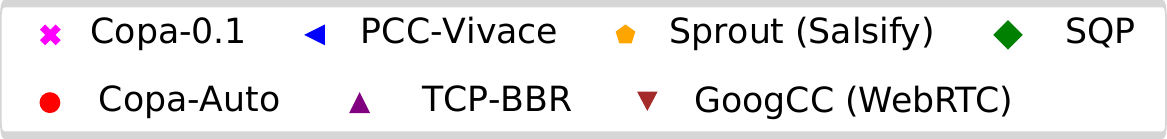}\\
    \subcaptionbox{Send rate timeseries.\label{fig:appendix:sqp_shortvar_timeseries}}{
    \includegraphics[width=0.7\columnwidth]{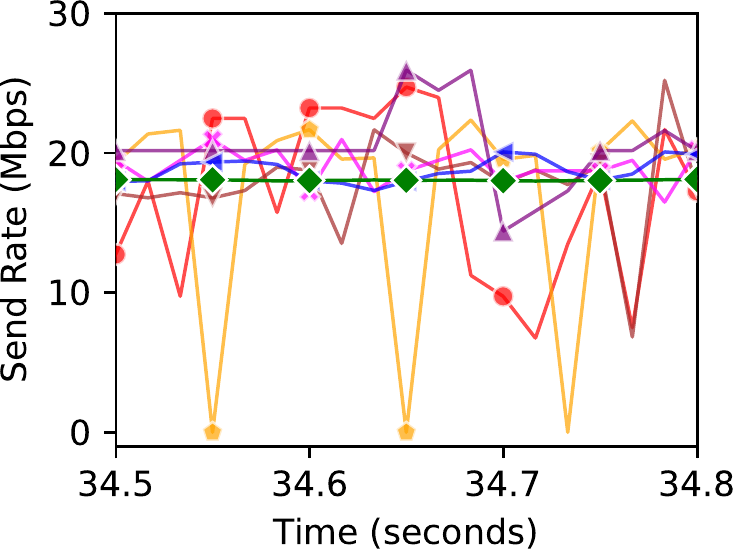}
    }%
    \hfill
    \subcaptionbox{Packet delay timeseries.\label{fig:appendix:sqp_shortvar_delay}}{
    \includegraphics[width=0.7\columnwidth]{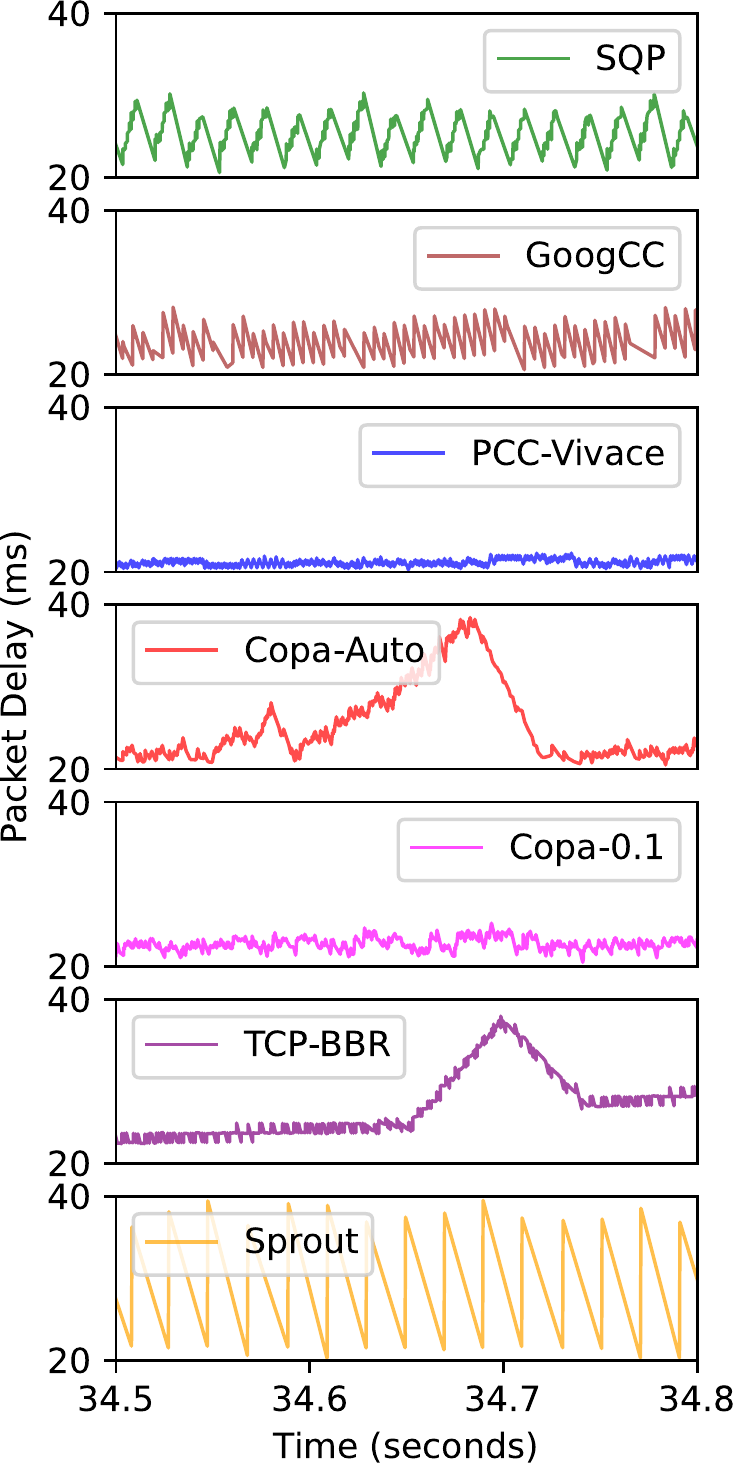}
    }%
    \caption{Short-timescale throughput and delay behavior on a 20 Mbps link (link bandwidth shown as a shaded light blue line).}
    \label{fig:appendix:experiment_shortvar_timeseries}
\end{figure}
\label{sec:appendix:shortvar}
In Figure~\ref{fig:appendix:experiment_shortvar_timeseries}, we show the short-term throughput and delay behavior of \sqp and other various CCAs, over a period of 0.5 seconds (see \S~\ref{sec:background:short_var}).
Figure~\ref{fig:appendix:sqp_shortvar_timeseries} shows the transmission rate for each frame period (16.66 ms at 60 FPS). \sqp's transmission rate is very stable, and does not vary at all across multiple frames.
Figure~\ref{fig:appendix:sqp_shortvar_delay} shows the packet delay for various CCAs over 0.5 seconds.
Since \sqp transmits each packet as a short (paced) burst, it causes queuing at sub-frame timescales, but since \sqp does not use more than 90\% of the link (due to $T=0.9$, \S~\ref{sec:pacing_target}), there is no queue buildup that occurs across frames.
While sprout's probing looks similar, the queuing caused by Sprout is much higher.
We note that \sprout's dips in throughput may be caused by the fact that the burst frequency of the \sprout sender used in our test is 50 cycles per second. This may not be a factor for video streaming if the burst frequency matches the video frame rate. \sprout's inadequacy for low-latency interactive streaming applications primarily stems from its inability to achieve sufficient bandwidth when competing with other queue-building flows.

\subsection{Impact of Feedback Delay}
\begin{figure*}[htbp]
    \centering
    \includegraphics[width=0.9\textwidth]{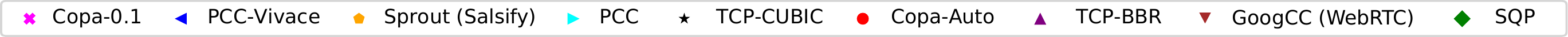}\\
    \subcaptionbox{Average delay after the link rate drops for various RTTs.\label{fig:eval:rtt_link_var}}{\includegraphics[width=0.23\textwidth]{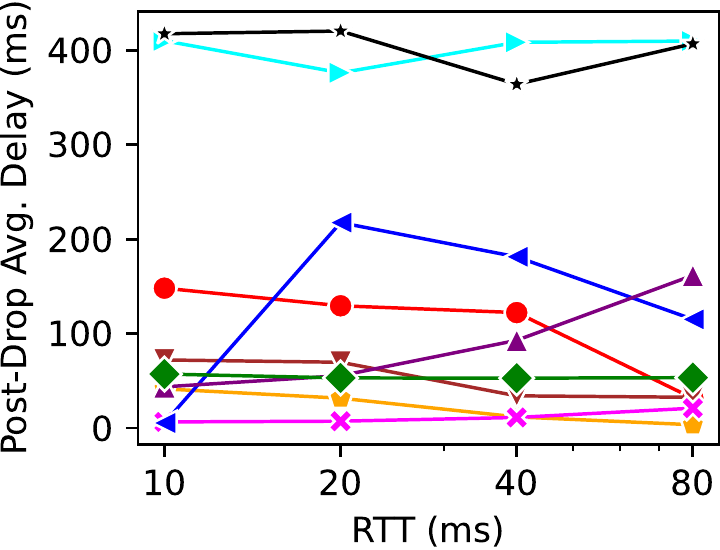}}%
    \hfill
    \subcaptionbox{Average link utilization for different RTTs. Sprout has low utilization on higher RTT links, and thus, lower delay.\label{fig:appendix:rtt_delay_tput}}{
    \includegraphics[width=0.23\textwidth]{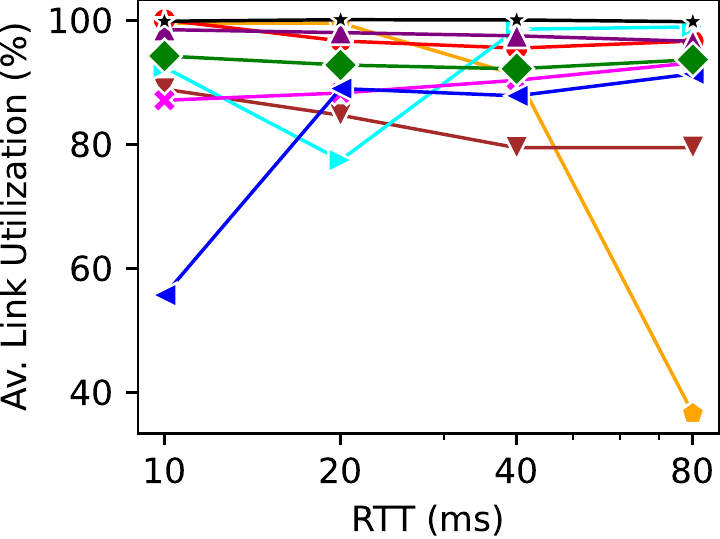}
    }%
    \hfill
    \subcaptionbox{Packet delay timeseries for 10ms RTT link. Copa-Auto runs in competitive mode on low RTT links (10ms).\label{fig:appendix:rtt_timeseries}}{
    \includegraphics[width=0.23\textwidth]{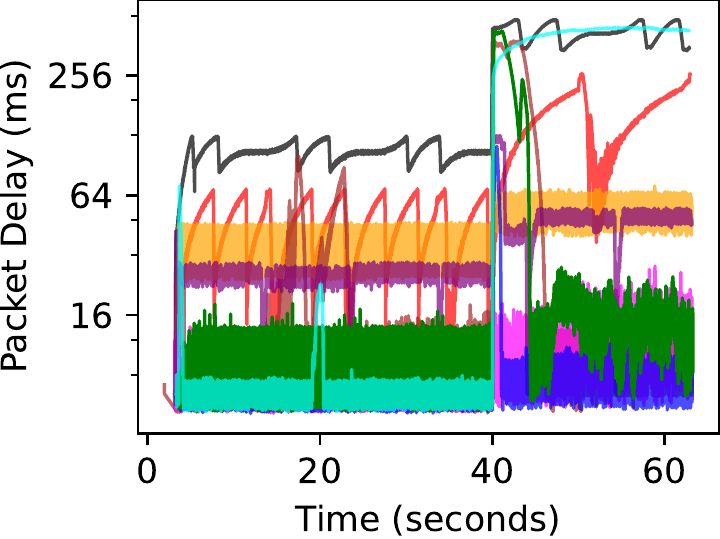}
    }
    \caption{Impact of link RTT on throughput and delay, where link changes from 20 Mbps to 5 Mbps at T=40s.}
    \label{fig:appendix:rtt_dynamics}
\end{figure*}

Since \sqp receives the frame delivery statistics at the sender after 1-RTT, it is important to evaluate the impact of delayed feedback on \sqp's dynamics. 
Figure~\ref{fig:eval:rtt_link_var} shows the average delay, and Figure~\ref{fig:appendix:rtt_delay_tput} shows the average throughput after a 20 Mbps link steps down to 5 Mbps, for various baseline network RTT values.
The link is run at 20 Mbps for 40 seconds, following which the link is run at 5 Mbps for an additional 20 seconds. Figure~\ref{fig:eval:rtt_link_var} shows the average delay for the last 20 seconds, when the link is operating at 5 Mbps.
\sqp's performance is consistent across the entire range, even though the feedback is delayed, and can be attributed to the fact that \sqp uses a larger window for $\Delta_{min}$ on higher RTT links. 
\sprout and Copa-Auto have lower delay on higher RTT links, but for different reasons: Sprout's link utilization drops sharply as the link RTT increases from 40 to 80 ms (Figure~\ref{fig:appendix:rtt_delay_tput}), and hence, it's delay is lower, whereas
Copa-Auto incorrectly switches to competitive mode on low RTT links, causing very high delays (Figure~\ref{fig:appendix:rtt_timeseries} shows the delay timeseries for a 10ms RTT link).
\pcc-Vivace can only maintain low delay across a 10ms RTT link, and \pcc is unable to drain the queuing caused after the link rate drops in all cases.
}

\subsection{Fairness}
\label{sec:eval:fairness}

\begin{figure}
    \centering
    \includegraphics[width=\columnwidth]{figures/legend_sigcomm_sqp.pdf}
    \subcaptionbox{Throughput-delay performance of 10 flows sharing a \SI{60}{Mbps} bottleneck link.\label{fig:eval:10flow:60mbit_throughput_delay}}{
    \includegraphics[width=0.47\columnwidth]{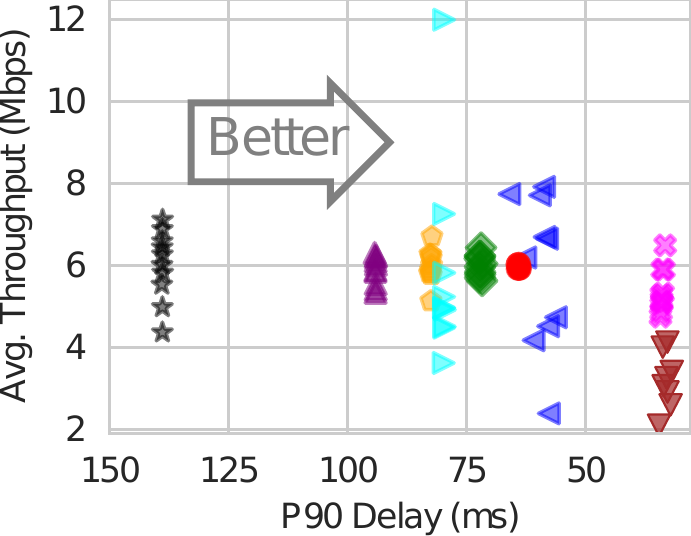}
    }%
    \hfill
    \subcaptionbox{Bitrate-frame delay performance of 10 flows sharing a \SI{60}{Mbps} bottleneck link.\label{fig:eval:10flow:60mbit_bitrate_delay}}{
    \includegraphics[width=0.46\columnwidth]{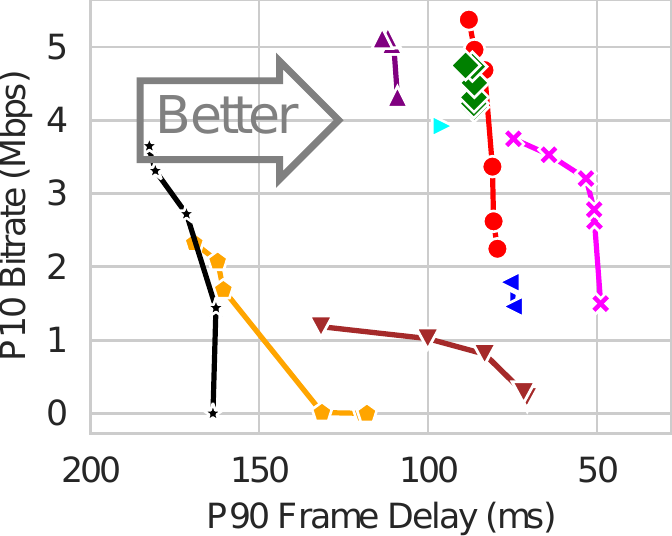}
    }
    \caption{Fairness results with 10 flows sharing a bottleneck. SQP achieves a fair share of throughput on average, and at smaller time-scales. CCAs like Sprout, WebRTC, and Copa become excessively bursty at smaller time-scales.}
\end{figure}

\begin{figure}
    \centering
    \includegraphics[width=\columnwidth]{figures/legend_sigcomm_sqp.pdf}
    \subcaptionbox{Jain's fairness index over time for homogeneous flows entering and exiting the bottleneck link.\label{fig:eval:fairness:fairness_30frames}}{
    \includegraphics[width=0.47\columnwidth]{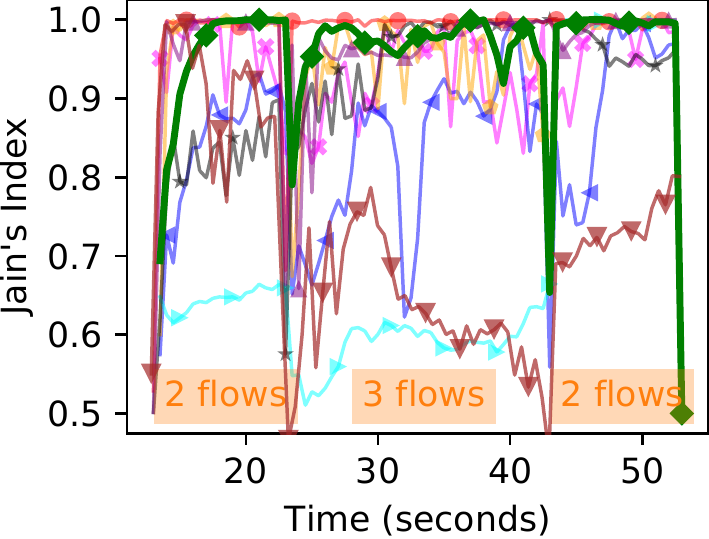}
    }%
    \hfill
    \subcaptionbox{P10 fairness (500 ms windows) for flows with different RTTs. First flow RTT = 20 ms. \label{fig:eval:fairness:rtt_fairness}}{
    \includegraphics[width=0.47\columnwidth]{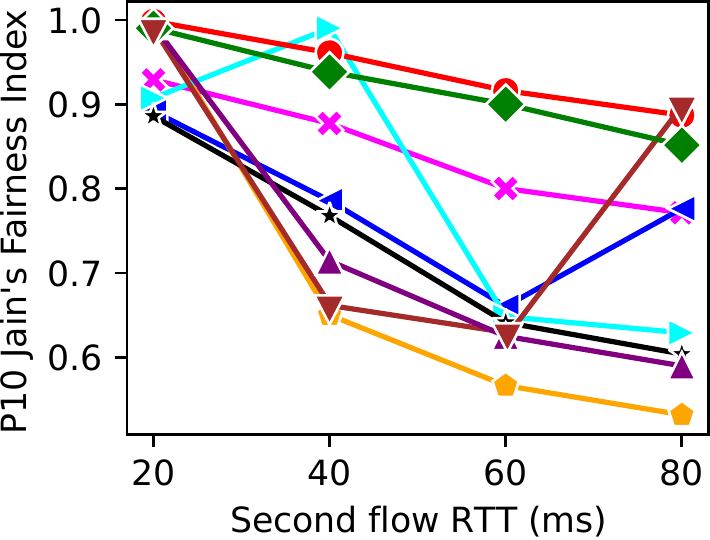}
    }
    \caption{Dynamic fairness and RTT fairness comparison. SQP quickly converges to fairness, and has good RTT fairness. }
\end{figure}

The first experiment evaluates the performance of 10 homogeneous flows sharing a \SI{60}{Mbps} bottleneck link, with a link RTT of \SI{40}{\milli\second}.
Figure~\ref{fig:eval:10flow:60mbit_throughput_delay} compares the average throughput and the P90 one-way packet delay for each flow.
The ideal behavior is that each flow achieves exactly 6 Mbps and low delay, ie. the points should be clustered at the 6 Mbps line and be towards the right in the plot.
\sqp flows\footnote{Inter-frame timing jitter enabled (\S~\ref{sec:bw_samp_analysis})} achieve equal share of the link, with lower P90 one-way packet delay compared to \sprout, \pcc, \bbr and \cubic (75 ms).
While \bbr, and both versions of \copa achieve good fairness with respect to the average throughput for the full experiment duration, neither version of \pcc is able to do so.
While \webrtc has very low P90 packet delay, the flows cumulatively underutilize the link and do not achieve fairness.
Figure~\ref{fig:eval:10flow:60mbit_bitrate_delay} compares the streaming performance of the algorithms by plotting the P10 bitrate and the P90 frame delay for different bitrate estimation windows ranging from 1 frame to 32 frames in multiplicative steps of 2 (\S~\ref{sec:eval:metrics}).
The streaming performance of \copa-0.1, \sprout, \cubic and \webrtc are significantly worse than their average throughput and packet delay due to bursty transmissions when multiple flows share the bottleneck link.

In the second experiment, we evaluate dynamic fairness as flows join and leave the network.
Flows 2 and 3 start \SI{10}{\second} and \SI{20}{\second} after the first flow respectively, and stop at \SI{40}{\second} and \SI{50}{\second} respectively.
Figure~\ref{fig:eval:fairness:fairness_30frames} plots the Jain fairness index~\cite{jain1984quantitative} computed over 500 ms windows versus time.
\sqp converges rapidly to the fair share, whereas both versions of \pcc, \copa-0.1 and \webrtc cannot reliably achieve fairness at these time scales.

\sqp also demonstrates good fairness across flows with different RTTs.
We evaluated steady-state fairness among flows that share the same bottleneck, but have different network RTTs.
In Figure~\ref{fig:eval:fairness:rtt_fairness}, we plot the P10 fairness (using Jain's fairness index) across windowed 500 ms intervals.
When two flows have the same RTT, \sqp, \copa-Auto, TCP-\bbr and \sprout achieve perfect fairness.
As the RTT of the second flow increases, \sqp and \copa-Auto are able to maintain reasonable throughput fairness but the fairness degrades rapidly in the case of TCP-\bbr, \cubic and \sprout as the RTT of one flow increases.
The slight drop in fairness in the case of \sqp is because the flow with the higher RTT achieves lower throughput since its minimum one way delay window size is larger. \pcc, \vivace, and \webrtc also achieve low fairness for flows with different RTTs and do not demonstrate any particular pattern as the RTT difference between the flows increases.

\subsection{\sqp Video Codec Integration}
\label{sec:encoder_variation}

\begin{figure}
    \centering
    \includegraphics[width=0.8\columnwidth]{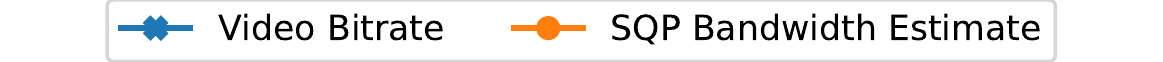}
\subcaptionbox{\sqp in isolation.\label{fig:applim_isolated}}{\includegraphics[width=0.49\columnwidth]{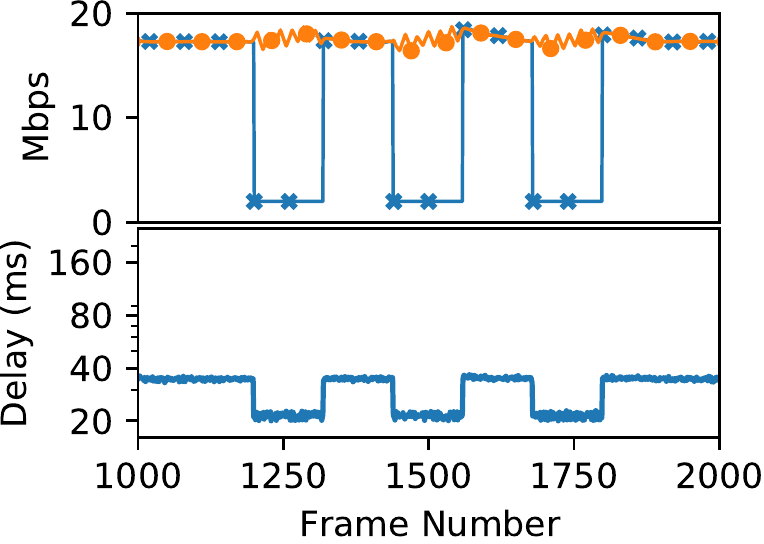}}%
\hfill
    \subcaptionbox{Competing with \cubic. \label{fig:applim_cubic}}{\includegraphics[width=0.49\columnwidth]{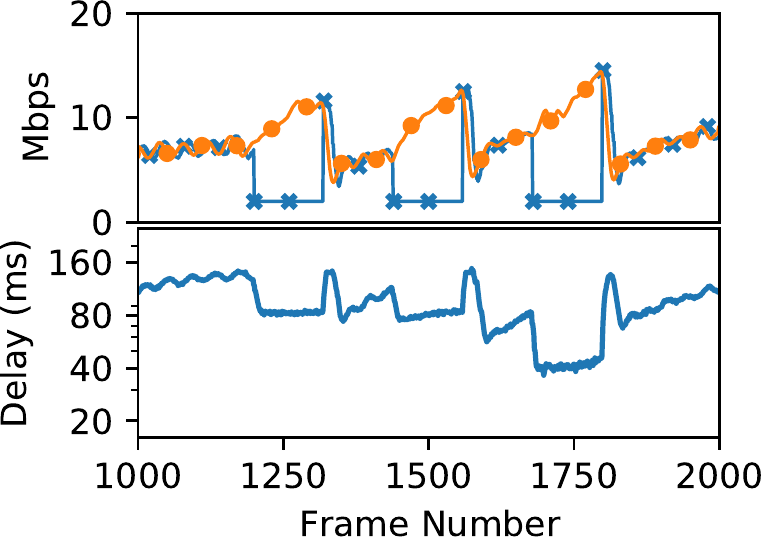}}%
    \caption{\sqp's performance when application-limited.}
    \label{fig:enc_undershoot}
\end{figure}

We evaluated \sqp's bandwidth estimation in a scenario where the video bitrate is significantly lower than the bandwidth estimate. 
We tested \sqp by artificially limiting the video bitrate on a 20 Mbps, 40 ms RTT link with 120 ms of bottleneck buffer. 
The encoder bitrate is artifically capped to 2 Mbps for three 2-second intervals.
In Figure~\ref{fig:applim_isolated}, \sqp maintains a high bandwidth estimate, which is appropriate since \sqp is the only flow on the link.
\sqp also obtains a reasonable estimate of the link bandwidth under application-limited scenarios when competing with other flows.
Figure~\ref{fig:applim_cubic} shows \sqp's bandwidth estimate when the video bitrate is lower than the target bitrate and \sqp is competing with a Cubic flow. When the video bitrate is lower than the target, \sqp is able to maintain a high bandwidth estimate, which demonstrates that \sqp is able to maintain a high bandwidth estimate without requiring additional padding data.
This allows \sqp to quickly start utilizing its share when the video bitrate is no longer limited (matches the target bitrate), instead of acquiring its throughput share from scratch, which would take much longer.
These experiments demonstrate that padding bits are not necessary for \sqp to achieve good link utilization.

The generated video bitrate can also overshoot the requested target bitrate.
In such scenarios, it is typically the encoder's responsibility to make sure that the average video bitrate matches the requested target bitrate, although \sqp can handle and recover from occasional frames size overshoots since they would cause subsequent bandwidth samples to be lower.
Persistent overshoot can occur in very complex scenes when the target bitrate is low.
In such cases, the application must take corrective actions that include reducing the frame rate or changing the video resolution.
Salsify~\cite{fouladi2018salsify} proposes encoding each frame at two distinct bitrates, choosing the most appropriate size just before transmission.
While \sqp can serve as a viable replacement for \sprout in Salsify, in Appendix~\ref{sec:appendix:encoderovershootundershoot}, we show that modern encoders like NVENC~\cite{patait2016nvenc} have good rate control mechanisms that avoid overshoot and can consistently match the requested target bitrate.

\section{Real-World Performance}
\label{sec:real_world_copa_ab}
\begin{figure}
\centering
    \subcaptionbox{Wi-Fi performance.\label{figures/eval/real_world/wifi}}[0.48\columnwidth]{\includegraphics[width=0.5\columnwidth]{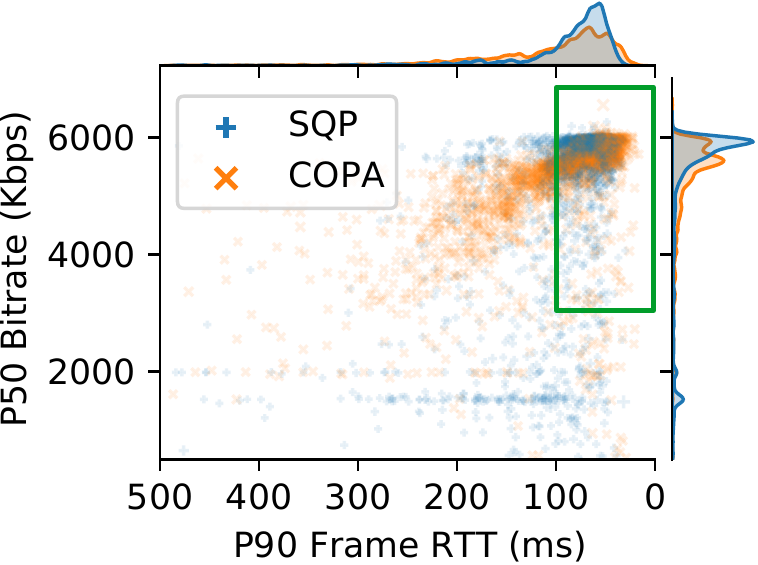}}%
    \hfill
     \subcaptionbox{LTE performance.\label{figures/eval/real_world/lte}}[0.48\columnwidth]{\includegraphics[width=0.5\columnwidth]{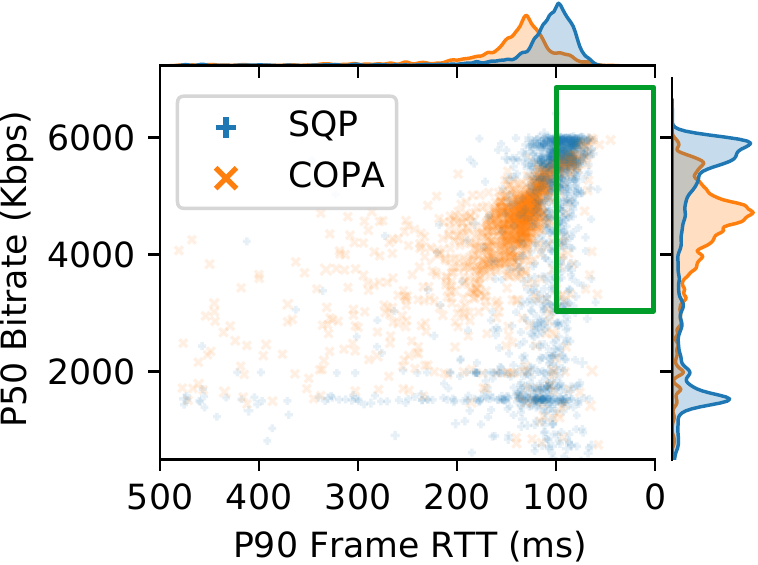}}%
     \caption{Real world A/B testing of \sqp and \copa-0.1.}
     \label{fig:eval:real_world_AB}
\end{figure}

\balance
To evaluate \sqp's performance in the real world, we deployed \sqp in \company's AR streaming platform.
We also deployed \copa-0.1 (without mode switching) on the same platform by adapting the MVFST implementation of \copa~\cite{mvfst-copa} and performed A/B testing, comparing the performance of the two algorithms.
We chose \copa-0.1 since it consistently maintained low delay compared to other CCAs (e.g. Sprout (Salsify) has very high delays) on emulated tests, and has been demonstrated to work well for low-latency live video in a production environment~\cite{garg-2020}~\footnote{In addition, Salsify's custom software encoder cannot sustain the frame rates required for low-latency interactive streaming applications}.
For \copa, we use $\frac{\mathrm{CWND}}{\mathrm{sRTT}}$ to set the encoder bitrate, and reduce the bitrate by $\frac{Q_{sender}}{D}$ when sender-side queuing occurs ( $Q_{sender}=$ pending bytes from previous frames, $D = 200 \si{\milli\second}$ is a smoothing factor), gradually reducing the sender-side queue over a period of 200 ms.
We ran the experiment for 2 weeks and obtained data for approximately 2400 Wi-Fi sessions and 1600 LTE sessions for each algorithm.
Figure~\ref{fig:eval:real_world_AB} shows the scatter plots of the median bitrate and the P90 frame RTT (fRTT; send start to notification of delivery) in addition to the separate distributions for each metric.
64 \sqp and 105 \copa sessions over LTE, and 36 \sqp and 52 \copa sessions over Wi-Fi had a P90 fRTT higher than 500 ms, and these are not shown in the figure.

71\% of \sqp sessions over Wi-Fi had good performance (bitrate $>$ 3 Mbps, fRTT $<$ 100 ms), compared to 56\% of \copa-0.1 sessions.
On LTE links, 36\% of \sqp sessions  had good performance, compared to 9\% of Copa sessions.
Across all the sessions, fRTT was less than 100 ms for 64\% of \sqp sessions and only 39\% of \copa sessions. These regions are highlighted with green boxes in Figure~\ref{fig:eval:real_world_AB}.

\sqp achieves lower frame delay compared to \copa across both Wi-Fi and LTE.
\sqp on Wi-Fi also achieves higher bitrate compared to \copa.
On LTE connections, \sqp demonstrates a bi-modal distribution of the bitrate, with a significant number of sessions being stuck at a low bitrate despite having a low RTT.
We believe \sqp gets stuck at a low bandwidth estimate due to a combination of noisy links, a low bandwidth estimate and encoder undershoot, although this needs to be investigated further (Eq.~\ref{eq:bw_samp_undershoot} was not used).
On the other hand, the bitrates for \copa sessions over LTE are more evenly distributed, but also incur higher delays compared to \sqp.

Our results from emulation and real-world experiments demonstrate that \sqp can efficiently utilize wireless links with time-varying bandwidth and simultaneously maintain low end-to-end frame delay, making it suitable for  wireless AR streaming and cloud gaming applications.

\vspace{-0.1in}
\section{Conclusion}
\label{sec:conclusion}
In this paper, we have presented the design, evaluation, and results from real-world deployment of \sqp, a congestion control algorithm designed for low-latency interactive streaming applications.
\sqp is designed specifically for low-latency interactive video streaming, and makes key application-specific trade-offs in order to achieve its performance goals.
\sqp's novel approach for congestion control enables it to maintain low queuing delay and high utilization on dynamic links, and also achieve high throughput in the presence of queue-building cross traffic like \cubic and \bbr, without the caveats of explicit mode-switching techniques.

\clearpage

\bibliographystyle{IEEEtran}
\bibliography{main}

\clearpage
\begin{appendices}

\section{Video Encoder Overshoot/Undershoot}
\label{sec:appendix:encoderovershootundershoot}

\srini{start with a reference to salsify -- past studies have focused on the mismatch between encoder output and requested rate. SQP doesn't have an explicit mechanism to handle this.... To see if this is significant issue, ....}\dd{PTAL}
Previous systems like Salsify have focused on the challenges of integrating a video codec with the congestion control algorithm due to mismatches that can occur between encoder output and the requested rate.
Salsify proposes a custom video encoder that encodes video frames at two different target bitrates, and chooses the largest frame whose transmission will not exceed the available network bandwidth.
Since \sqp does not have an explicit mechanism to handle overshoots (the bandwidth estimate is penalized in \sqp due to transient queuing, but this may not be enough), \sqp must rely on the accuracy of the video encoder's rate control to produce accurate frame sizes.
We evaluate the rate-control accuracy (whether the encoder is able to produce frames close to the requested target size) of a modern commercial off-the-shelf hardware video encoder, NVENC~\cite{patait2016nvenc}, that natively supports real-time interactive video streaming workloads.
We use the videos from a cloud gaming video data set (CGVDS~\cite{cgvds}, 1080p@60FPS, 30s duration) for evaluating the encoder's rate-control accuracy.
We encode the video by randomly changing the target bitrate for each frame, where the bitrate is sampled from a uniform distribution between $[2\;\mathrm{Mbps},20\;\mathrm{Mbps}]$.
The NVENC settings were chosen according to the values specified in the official NVENC documentation~\cite{nvidia-settings} for low-latency video streaming.

\begin{figure*}[htbp]
    \centering
    \subcaptionbox{Videos used for testing (CGVDS)\label{fig:rc_dataset}}{
    \includegraphics[width=0.35\columnwidth]{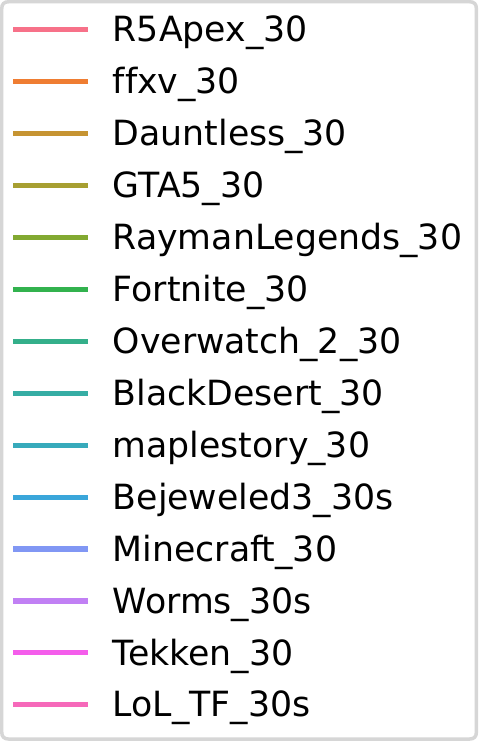}}%
    \hfill
    \subcaptionbox{Rate-control accuracy for different target bitrates\label{fig:rc_accuracy_bitratebin}}{
    \includegraphics[width=0.3\textwidth]{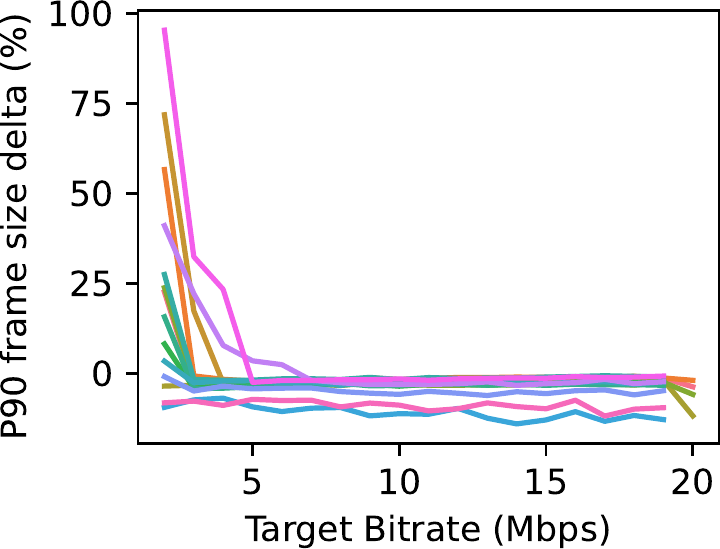}}%
    \hfill
\subcaptionbox{Rate-control accuracy for 2-4 Mbps target bitrate range. \label{fig:rc_accuracy_2mbps}}{\includegraphics[width=0.3\textwidth]{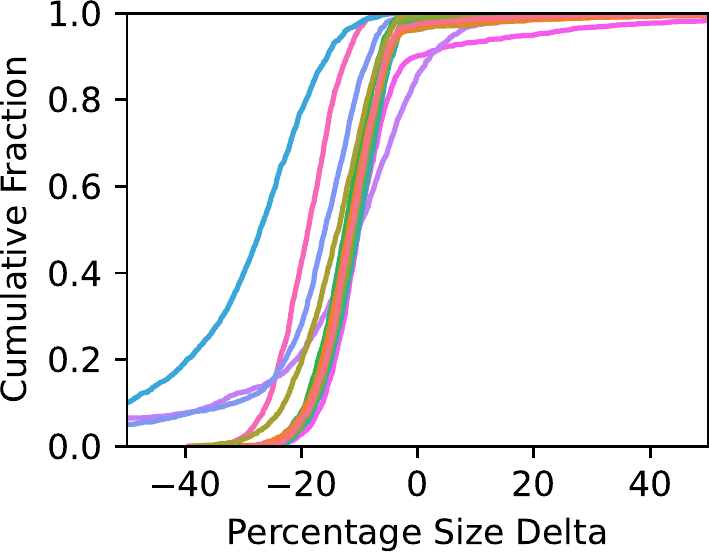}}
    \caption{
    Rate-control accuracy of the NVENC encoder, tested in low-latency configuration.
    }
\end{figure*}

The results are shown in Figure~\ref{fig:rc_accuracy_bitratebin}. 
The X-axis is the requested target bitrate, and the Y-axis plots the 90th percentile percentage frame size delta ($100 \cdot \frac{\mathrm{actual} - \mathrm{requested}}{\mathrm{requested}}$) for that bitrate for various videos.
For bitrates including 5 Mbps and above, the encoder does an excellent job of keeping the video bitrate under the target bitrate. 
Some overshoot occurs at lower rates ($\le 4;\mathrm{Mbps}$).
In Figure~\ref{fig:rc_accuracy_2mbps}, we plot the CDF of the frame size delta for target bitrates 4 Mbps and under. In this case, we observe that the encoder is still able to do a good job, and only occasionally overshoots the target bitrate. Worms\_30s and LoL\_TF\_30s have the highest fraction of frames that overshoot the target bitrate, but this fraction is also low (around 15\%).

Our conclusion is that with modern encoders like NVENC, in practive, video bitrate overshoots only happen at the lowest bitrates, and are otherwise not a major concern.
Overshoot can be attributed to there just being too much data to encode in the video, and not rate control accuracy specifically, and solutions may include reducing the video resolution in order to accommodate the lower bitrates.
While techniques like Salsify~\cite{fouladi2018salsify} are useful for low bitrate operation, where accurately controlling frame sizes is critical, and challenging, modern hardware-based codecs already do a good job at controlling video bitrate overshoot at bitrates commonly used for applications like cloud gaming and AR streaming.

Video bitrate undershoots are more common, and \sqp is able to handle these scenarios well (\S~\ref{sec:encoder_variation}, , \S~\ref{sec:encoder_variation}).

\section{GoogCC startup behavior}
\label{sec:appendix:webrtc_startup}
\begin{figure*}
    \centering
    \includegraphics[width=0.65\columnwidth]{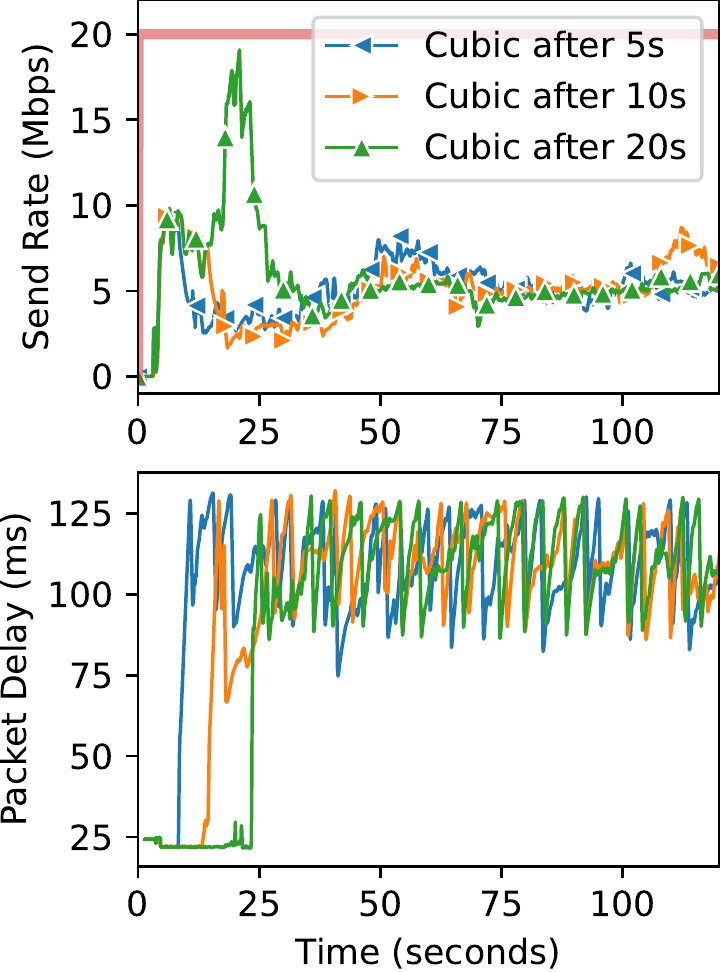}
    \caption{GoogCC's slower startup affects short-term throughput when competing with \cubic.}
    \label{fig:webrtc_Startup}
\end{figure*}
In \S~\ref{sec:eval:cross_traffic}, we start the competing \cubic flow after 10 seconds, but GoogCC does not converge to its maximum throughput at steady state by that point.
Thus, the throughput is slightly lower than if it had reached steady state.
Figure~\ref{fig:webrtc_Startup} shows the throughput and delay for the same experiment, but we vary the start time of the competing flow between 5 and 20 seconds.
While GoogCC's throughput is high for a short period of time right after the \cubic flow starts, the GoogCC flows eventually converge to the same rate.

\end{appendices}

\end{document}